\documentclass{aa}  
\usepackage{txfonts}
\usepackage{lscape}
\usepackage{xcolor}
\usepackage{graphicx}
\usepackage{txfonts}
\usepackage[colorlinks=true,allcolors=blue]{hyperref}

\newcommand{\nii}{[N{\sc{ii}}]}
\newcommand{\lognii}{$\log$([N{\sc{ii}}]/H$\alpha$)}
\newcommand{\logoii}{$\log$([O{\sc{iii}}]/H$\beta$)}

\newcommand{\oiid}{[O{\sc{ii}}]$\lambda\lambda$3727,3729}

\newcommand{\cii}{[C{\sc{ii}}]}
\newcommand{\oiii}{[O{\sc{iii}}]}
\newcommand{\ha}{H$\alpha$}
\newcommand{\lya}{Ly$\alpha$}

\newcommand{\hg}{H$\gamma$}

\newcommand{\hb}{H$\beta$}
\newcommand{\siid}{[S{\sc{ii}}]$\lambda\lambda6717,6731$}
\newcommand{\sii}{[S{\sc{ii}}]}
\newcommand{\oi}{[O{\sc{i}}]$\lambda6300$}

\newcommand{\hei}{He{\sc{i}}$\lambda5875$}

\newcommand{\neiii}{[Ne{\sc{iii}}]$\lambda3870$}

\newcommand{\sfr}{M${_\odot}$ yr$^{-1}$}        
\newcommand{\msun}{M${_\odot}$}

\newcommand{\kms}{${\rm km~s^{-1}}$}

\usepackage{xspace} 
\let\oldAA\AA
\renewcommand{\AA}{\text{\oldAA}\xspace}

\begin{document}

   \title{GA-NIFS: Multiphase analysis of a star-forming galaxy at $z\sim5.5$ }

   \subtitle{}
   \authorrunning{E. Parlanti et al.}

   \author{Eleonora Parlanti \thanks{E-mail: \href{mailto:eleonora.parlanti@sns.it}{eleonora.parlanti@sns.it}}
  \inst{\ref{inst:SNS}} 
    \and  
   Stefano Carniani  \inst{\ref{inst:SNS}}
\and 
Giacomo Venturi \inst{\ref{inst:SNS}}
\and
Rodrigo Herrera-Camus \inst{\ref{inst:concepcion}}
\and
Santiago Arribas \inst{\ref{inst:centroastrob}}
\and
Andrew J. Bunker \inst{\ref{inst:oxford}}
\and
Stéphane Charlot\inst{\ref{inst:paris}}
\and 
Francesco D'Eugenio  \inst{\ref{inst:kavli}, \ref{inst:cavendish}}
\and 
Roberto Maiolino  \inst{\ref{inst:kavli}, \ref{inst:cavendish}, \ref{inst:londoncollege}}
\and 
Michele Perna \inst{\ref{inst:centroastrob}}
\and
Hannah \"Ubler \inst{\ref{inst:kavli}, \ref{inst:cavendish}}
\and
Torsten Böker \inst{\ref{inst:esa}}
\and
Giovanni Cresci \inst{\ref{inst:arcetri}}
\and 
Mirko Curti \inst{\ref{inst:eso}}
\and
Gareth C. Jones \inst{\ref{inst:oxford}}
\and
Isabella Lamperti\inst{\ref{inst:dipflorence}, \ref{inst:arcetri},\ref{inst:centroastrob}}
\and 
Pablo G. P\'erez-Gonz\'alez \inst{\ref{inst:centroastrob}}
\and
Bruno Rodr\'iguez Del Pino\inst{\ref{inst:centroastrob}}
\and
Sandra Zamora \inst{\ref{inst:SNS}}}
   \institute{Scuola Normale Superiore, Piazza dei Cavalieri 7, I-56126 Pisa, Italy \label{inst:SNS} 
   \and
   Departamento de Astronomía, Universidad de Concepción, Barrio Universitario, Concepción, Chile \label{inst:concepcion}
   \and 
    Centro de Astrobiolog\'{\i}a (CAB), CSIC-INTA, Ctra. de Ajalvir km 4, Torrej\'on de Ardoz, E-28850, Madrid, Spain \label{inst:centroastrob} 
    \and
    University of Oxford, Department of Physics, Denys Wilkinson Building, Keble Road, Oxford OX13RH, United Kingdom \label{inst:oxford}
    \and
    Sorbonne Universit\'e, CNRS, UMR 7095, Institut d'Astrophysique de Paris, 98 bis bd Arago, 75014 Paris, France \label{inst:paris}
    \and
    Kavli Institute for Cosmology, University of Cambridge, Madingley Road, Cambridge, CB3 0HA, UK \label{inst:kavli} 
    \and
    Cavendish Laboratory - Astrophysics Group, University of Cambridge, 19 JJ Thomson Avenue, Cambridge, CB3 0HE, UK \label{inst:cavendish} 
    \and 
    Department of Physics and Astronomy, University College London, Gower Street, London WC1E 6BT, UK \label{inst:londoncollege} 
    \and
    European Space Agency, c/o STScI, 3700 San Martin Drive, Baltimore, MD 21218, USA \label{inst:esa}
    \and
INAF - Osservatorio Astrofisco di Arcetri, largo E. Fermi 5, 50127 Firenze, Italy\label{inst:arcetri}
    \and
European Southern Observatory, Karl-Schwarzschild-Strasse 2, 85748 Garching, Germany \label{inst:eso}
\and 
 Dipartimento di Fisica e Astronomia, Università di Firenze, Via G. Sansone 1, 50019, Sesto F.no (Firenze), Italy \label{inst:dipflorence}
 }
   \date{}
 
  \abstract
   {
   In this study, we present a detailed multiphase analysis of HZ4, a main-sequence star-forming galaxy at $z \sim 5.5$, known for being a turbulent rotating disk and having a detection of a \cii\ outflow in the ALMA observations. 
   We exploited JWST/NIRSpec observations in the integral field spectroscopy mode with low- and high-spectral resolution which allow us, for the first time, to spatially resolve the rest-frame UV and optical emission of the galaxy to investigate the galaxy properties.
   In particular, the high-resolution dataset allowed us to study the kinematics of the ionized gas phase, and the conditions of the interstellar medium, such as the excitation mechanism, dust attenuation, and metallicity.
   The lower spectral-resolution observations allowed us to study the continuum emission and infer the stellar populations' ages and properties.
   Our findings suggest that HZ4 is a galaxy merger rather than a rotating disk as previously inferred from lower-resolution \cii\ data. 
   The merger is associated with an extended broad, blueshifted emission, potentially indicative of an outflow originating from a region of intense star formation and extending up to 4 kpc.    
   In light of these new observations, we reanalyzed the ALMA data to compare the multiphase gas properties. 
    If we interpret the broad components seen in \cii\ and \oiii$\lambda$5007\AA as outflows, the neutral and ionized components are co-spatial, and the mass loading factor of the ionized phase is significantly lower than that of the neutral phase, aligning with trends observed in multiphase systems at lower redshifts. Nonetheless, additional observations and larger statistical samples are essential to determine the role of mergers and outflows in the early Universe and to clarify the origin of the broad emission components observed in this system.
   }


   \maketitle
%

\section{Introduction}

\label{sec:introduction}

 Several theoretical models suggest that galactic outflows driven by star formation (SF) and active galactic nuclei (AGNs) are crucial to explain the lack of galaxies in both the high- and low-mass ends of the galaxy mass function \citep[e.g.,][]{Fabian:2012, Naab:2017, Somerville:2015}, the inefficiency of galaxies in turning baryons into stars \citep{Dekel:1986, Behroozi:2019}, the metal enrichment of the circumgalactic medium (CGM) and the intergalactic medium (IGM) \citep{Oppenheimer:2006}, and the shape of the mass-metallicity relation \citep{Chisholm:2018}.
In particular, low-mass ($<10~{\rm M_\odot}$) galaxies are believed to self-regulate the buildup of their stellar mass through the action of SF-driven outflows \citep{Hopkins:2012}.
The energy and momentum injected by young massive stars in the shape of stellar winds, supernova (SN) explosions, and radiation pressure can provide feedback, by heating and expelling the gas from the galaxy \citep[e.g.,][]{Somerville:2015}.  

Local observations of low-mass galaxies have revealed that they also host accreting supermassive black holes at their centers \citep[e.g.,][]{Sartori:2015, Reines:2020}, indicating that SF activity can be regulated by AGN-driven outflows. Theoretical works and simulations have indeed pointed out that AGNs can boost the outflow temperature and velocity by two orders of magnitude \citep{Silk:2017, Dashyan:2018, Koudmani:2019}, which means that AGN-driven outflows may affect the CGM around low-mass galaxies by polluting it with metals and heating it. \cite{Koudmani:2021} have also shown that AGN feedback in low-mass galaxies may be more relevant at $z > 2$, when AGN are expected to be more active.  In conclusion, the primary mechanisms responsible for regulating star formation in low-mass galaxies are still unclear and debated. Theoretical models offer a wide range of solutions to this problem, and multiwavelength observations are needed to test them.

In the last ten years, observations and simulations have shown that galactic outflows include gas at different temperatures, densities, and states (molecular, neutral atomic, and ionized) \citep[e.g.,][]{Muratov:2015,Janssen:2016, Li:2017, Nelson:2019, Fluetsch:2019, Fluetsch:2021}. In particular,  four components have been identified: a) the hot  (T $\sim 10^{6-7}$~K) ionized outflowing gas, which emits in the X-rays; b) the warm ionized component (T $\sim 10^4$~K), which is usually detected as blueshifted broad wings in the brightest optical emission lines, such as \oiii$\lambda$5007\AA and H$\alpha$; c) the cool neutral atomic part (T $\sim 100$~K) that is usually observed via NaID $\lambda\lambda$5890,5896\AA\ in the local Universe and  \cii$\lambda$158$\mu$m\footnote{The fraction of \cii\ emission arising from ionized gas is only 1-30\% of the total emission \citep{Diaz-Santos:2017, Cicone:2018}} in high-$z$ galaxies; and d) the cold molecular component (T$\sim$ 10 K), traced usually by the broad wings of the CO molecular transitions or by P-Cygni profiles of molecular transitions (e.g., OH 119 $\mu$m and H$_2$0).
Aside from a few studies, the majority of the outflows are observed only through one tracer, making it difficult to estimate the global outflow properties such as the total outflow mass and mass loading factor, hence the effect on the host galaxy.

While at cosmic noon ($1\leq z \leq 3$) outflows are usually identified from broad, often blueshifted components of optical emission lines \citep[e.g., ][]{Coatman:2019,Forster:2019,Villarmartin:2020, Concas:2022, Cresci:2023}, at high redshift ($z>3$) outflow studies have focused on the far-infrared (FIR) emission lines. This is because optical emission lines are redshifted to near-infrared (NIR) wavelengths, which become hard to access from the ground. In particular, Atacama Large Millimetre Array (ALMA) observations have revealed outflow signatures in molecular transitions (P-Cygni profiles of OH 119 $\mu$m and broad CO and H$_2$O line wings; \citealt{Herreracamus:2019, Jones:2019, Butler:2023}) and the atomic carbon transition, \cii\ \citep{Gallerani:2018, Ginolfi:2020, Bischetti:2019, Herrera-Camus:2021, Solimano:2024_ALMA, Tripodi:2023}. 
However, the majority of these studies have mainly focused on AGN host galaxies, massive submillimeter galaxies, and lensed dusty star-forming galaxies \citep{Jones:2019, Spilker:2020}.
Evidence for SF-driven neutral outflows in galaxies in the early Universe has been observed for the first time by stacking the \cii\ emission of samples of galaxies at $z\sim4-5$  \citep{Gallerani:2018, Ginolfi:2020}. So far, broad \cii\ in the spectra of individual main-sequence galaxies has been reported only for two $z>4$ sources  \citep{Herrera-Camus:2021, Solimano:2024_ALMA}.

With the advent of the James Webb Space Telescope (JWST), we finally have access to the rest-frame optical emission lines of high-redshift galaxies, which allows us to study the outflows of high-z galaxies with the same tools developed for low-z ones.
This also gives us the possibility to simultaneously characterize up to two (or three, see above) outflow components in the same target.
Early studies with JWST have shown that a non-negligible fraction of star-forming galaxies -- about 20-30\% -- exhibit ionized outflows, even up to the cosmic dawn \citep{Carniani:2023, Zhang:2024, Calabro:2024}. Therefore, the synergy between ALMA and JWST promises to be a rich probe into the physical nature of high-$z$ outflows.

In this work, we focus on HZ4 (RA = 09h58m28.5s, Dec= +02d03m06.7s), which is the most distant star-forming galaxy with evidence of neutral outflows through the \cii\ emission line \citep{Herrera-Camus:2021}.
HZ4 was identified as a Lyman Break Galaxy (LBG) at $z \sim 5.5$ in the Cosmic Evolution Survey (COSMOS) field \citep{Scoville:2007} and was spectroscopically confirmed with Keck Deep Extragalactic Imaging Multi-Object Spectrograph (DEIMOS) observations \citep{Mallery:2012} which detected \lya\ emission  \citep{Cassata:2020, Lefevre:2020}.
It was then followed up with ALMA observations to target the \cii\ and dust continuum emission with low- (1.01\arcsec $\times$ 0.85\arcsec)  and high- (0.39\arcsec $\times$ 0.34\arcsec)  angular resolution \citep{Capak:2015, Bethermin:2020, Herrera-Camus:2021, Jones:2021}. In the ALPINE survey \citep{Lefevre:2020, Bethermin:2020}, HZ4 is dubbed DEIMOS\_COSMOS\_494057. 
The high-resolution ALMA \cii\ observations characterize the source as a prototypical high-$z$, turbulent, but rotating disk galaxy \citep{Herrera-Camus:2022, Parlanti:2023}. These spatially resolved observations also reveal the presence of a broader component in the spectrum of the \cii\ line emission which was interpreted as a neutral outflow \citep{Herrera-Camus:2021}. JWST observations HZ4 allow us to investigate the warm ionized gas phase and compare its properties with those of the cold gas.

This work is structured as follows: in Sect.~\ref{sec:observations} we describe the JWST and ALMA observations and data reduction, in Sect.~\ref{sec:integrated_analysis} we describe the analysis of the integrated rest-frame optical spectrum, in Sects. \ref{sec:analysis_resolved} and \ref{sec:ciianalysis} we analyze the JWST and ALMA data at a spaxel level. In Sect.~\ref{sec:SED}, we present the analysis of the spectral energy distribution (SED) and its results.
In Sect.~\ref{dsec:discussions}, we discuss possible interpretations of the broad emission lines that we observe.
In Sect.~\ref{sec:outflow} we speculate that the broad components that we observe are tracing a multiphase outflow, hence we analyze and discuss the outflow properties and their impact on the galaxy.
Finally, in Sect.~\ref{sec:conclusion} we draw our conclusions.
Throughout this work, we adopt the cosmological parameters from \citet{Planck:2015}: $H_0
= 67.7$ \kms\,Mpc$^{-1}$, $\Omega_m$ = 0.307, and $\Omega_\Lambda$= 0.691, giving 1\arcsec = 6.09 kpc at z = 5.54.


\section{Observations}

\label{sec:observations}

\subsection{JWST data}

HZ4 was observed with the NIRSpec instrument \citep{Jakobsen:2022} onboard JWST as part of the GA-NIFS \footnote{\hyperlink{https://ga-nifs.github.io/}{https://ga-nifs.github.io}} survey (program 1217 -- Integral Field Spectroscopy in COSMOS, PI: Nora Luetzgendorf) on 23 April 2023.
The galaxy was observed in the Integral field spectroscopy (IFS) mode \citep{Boker:2022} with both the G395H/F290LP and PRISM/CLEAR grating/filter combination with an eight-dither position of the medium cycling pattern.
The total integration time was $\sim 5$h for the high spectral resolution observations ($R \sim 1670 - 3600$, hereafter also referred as R2700),  which cover the wavelength range between 2.87--5.27 $\mu$m, and $\sim$ 1h for the prism observations ($R \sim 30-400$, hereafter also referred as R100), which span the wavelength range between 0.6--5.3 $\mu$m.

We retrieved the raw data from the Mikulski Archive for Space Telescopes (MAST) archive, then we processed them with the JWST pipeline version 1.11.1 and Calibration Reference Data System (CRDS) context \verb|jwst_1097.pmap|.
We run the standard steps of the public JWST pipeline to produce the final cubes. In the pipeline process, we used some customized code to improve the quality of the final data \citep[see][]{Perna:2023}.
In particular, during the first stage ``calwebb\_detector1'' that accounts for detector level corrections, we also corrected for the $1/f$ noise. At the end of the stage ``calwebb\_spec2'' that calibrates the data, we visually inspected the calibrated exposures, and we manually masked all regions affected by cosmic rays, persistences, and failed open shutters in the MSA mask. Moreover, we removed the outliers by calculating the derivative of the count rate along the dispersion direction and removing all data where the measurement was above the 98th and 95th percentage of the distribution for R2700 and R100, respectively \citep{Deugenio:2023}.
Then, we applied the third stage, ``calwebb\_spec3'' to combine each exposure and create the final datacubes with a spaxel size of 0.05\arcsec\ by using the ``drizzle'' weighting. 
Finally, we subtracted the background emission from each cube by removing the median spectrum computed in the target-free regions of the cubes.

\begin{figure}
    \centering
    \includegraphics[width=\hsize]{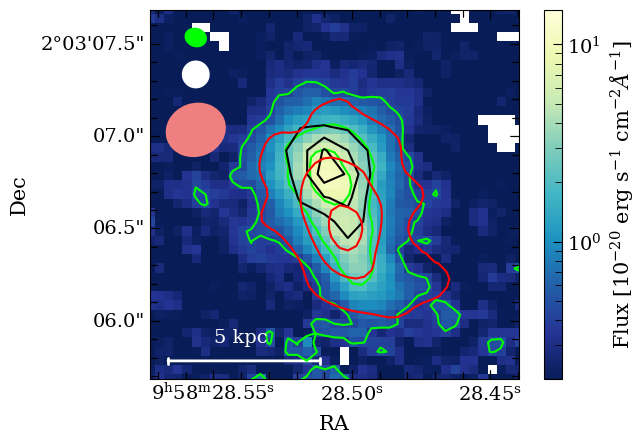}
    \caption{ Integrated \oiii$\lambda$5007\AA\ flux map of HZ4 from the JWST NIRSpec/IFS R2700 cube.  We show the 5, 25, and 75$\sigma$ of \oiii\ as green contours.
    Red contours illustrate the \cii\ emission from the ALMA Briggs-weighted (robust = 0.5) cube, at 5, 15, and 25$\sigma$, respectively. HST/WFC3 F160W image is reported in black contours.
    The pink, white, and green ellipses represent the ALMA beam size, the HST F160W point spread function (PSF), and NIRSpec IFS PSF at 3.2 $\mu$m estimated by \cite{Deugenio:2023}, respectively.}
    \label{fig:astrometry}
\end{figure}

\subsection{ALMA data}

The \cii\ raw data were retrieved from the ALMA archive (2012.1.00523.S
PI: Capak; 2017.1.00428.L, PI: Le Fevre; 2018.1.01605.S, PI: Herrera-Camus). The source was observed for $\sim 20$ min, $\sim 30$ min, and 4.7 h, respectively, in Band 7.
We used \verb'CASA' \citep{Mcmullin:2007} to combine these three observations and calibrate the \textit{uv} visibilities by using the script included in the dataset.
The measurement sets were then processed with CRISTAL’s reduction pipeline (see \citealt{posses:2024,Solimano:2024_ALMA, Villanueva:2024}; Herrera-Camus et al. in prep. for details). In brief, first, the continuum emission is subtracted in the visibility space using the task \verb'uvcontsub', the task \verb'tclean' is run with a Briggs weighting (robust parameter = 0.5 and 2), \verb'hogbom' deconvolver and a spaxel scale of 0.0445\arcsec. The resulting datacubes have a beam size of $0.33\arcsec \times0.29\arcsec$, and $0.45\arcsec \times0.41\arcsec$, respectively.
The cube with Briggs robust parameter = 2 (natural weighting) reaches a higher signal-to-noise ratio (S/N) at the expense of the angular resolution, while the cube with Briggs robust parameter = 0.5 is a compromise between the natural and the uniform (high angular resolution but lower S/N) weighting, providing us a good S/N and a higher angular resolution with respect to the natural weighting cube. All cubes have a channel width of 40 \kms.

ALMA observations also allow us to detect the FIR dust continuum emission at $\sim 160\, \mu$m in the galaxy rest frame. The continuum image was created with the task \verb'tclean' and a Briggs weighting with a robust parameter = 2 (natural weighting) to maximize the S/N. The final continuum image has a beam size of $0.46\arcsec \times 0.41 \arcsec$ and a sensitivity of 6~$\mu$Jy/beam.

\subsection{Astrometric registration}

We verified that the astrometric coordinates of JWST are aligned with those from ALMA. Since many other NIRSpec IFS observations 
showed an offset between the position indicated in the datacube header and the real position of the source on the sky \citep[e.g.,][]{Arribas:2023, Jones:2023, Lamperti:2023}, we assessed whether the astrometry of the JWST cubes of HZ4 was correct or not.

First, we retrieved the F160W images of HST/WFC3 (PID: 13641) from the MAST archive, with astrometry corrected a posteriori by using the GAIA eDR3 catalog with uncertainties on the order of 0.01\arcsec.
We then compared the image created by convolving the R100 cube with the F160W filter response and the image from the F160W filter from HST.
We fitted a 2D Gaussian profile to find the coordinates of the center both in the HST and JWST images.
The mismatch between the 2 images corresponds to a $\sim 0.10$\arcsec\ shift. We then shifted the JWST R100 cube to match the F160W astrometry.
We also ensured that the R2700 and R100 cubes were consistently aligned by collapsing the \oiii$\lambda$5007\AA\ emission line visible in both cubes. After verifying the alignment, we shifted the R2700 cube by the same amount as the R100 cube.
We assumed that the ALMA astrometry is correct, considering that the ALMA positional accuracy varies from 5\% to 20\% of the beam depending on the S/N, which corresponds to an accuracy of 0.015\arcsec -- 0.060\arcsec. The high S/N ensures that the ALMA astrometry is correct within 1 spaxel  (i.e., 0.05\arcsec).
Fig. \ref{fig:astrometry} shows the \oiii\ integrated emission along with the contours from the HST F160W image and the \cii\ integrated emission after the astrometric correction.


\section{Integrated analysis}

\begin{figure*}
    \centering
    \includegraphics[width=\hsize]{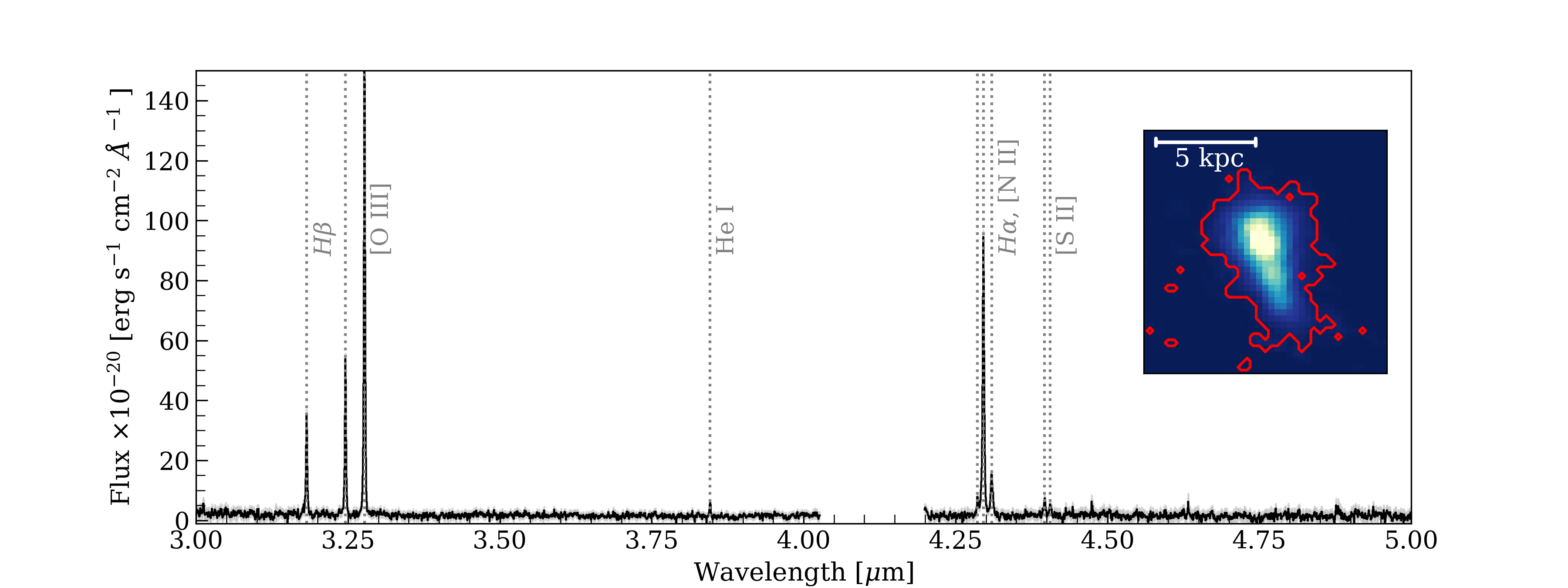}

    \includegraphics[width=\hsize]{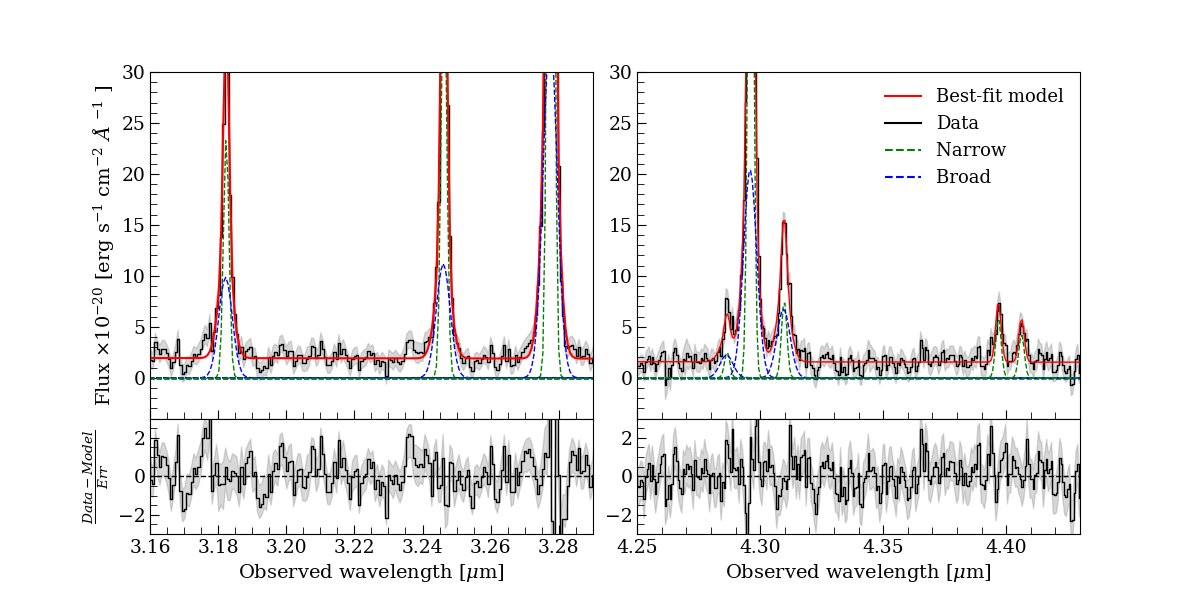}
    \caption{ High resolution spectrum of the source. Upper panel: Integrated spectrum of the source extracted from the regions with $S/N > 3$ in the wavelength range 3.27 -- 3.29 $\mu$m encompassing the \oiii$\lambda5007\AA$ emission. The extraction region is shown as the red contour in the inset panel.
    Bottom panels: Zoom-in on the integrated spectrum  (black) and our best fit (red). 
    The narrow and broad components are shown as the green and blue dashed lines, respectively.
     The \hb\ and \oiii\ complex is shown in the left panel, and the \ha, \nii, and \siid\ complex is shown in the right panel. The gray shaded areas mark the uncertainties on the data.}
    \label{fig:integrated_spectrum}
\end{figure*}

In this section, we describe the analysis of the R100 and R2700 NIRSpec IFS spatially integrated spectra of HZ4.  
The spatially resolved analysis is reported in Sect.~\ref{sec:analysis_resolved}.

\label{sec:integrated_analysis}

\subsection{Spectral fitting of the R2700 cube}
\label{sec:analysis_R2700}
In the top panel of Fig. \ref{fig:integrated_spectrum} we show the R2700 spectrum integrated over the regions of the cube with $S/N > 3$ in the wavelength range between 3.27 -- 3.29 $\mu$m, which encompasses the \oiii$\lambda 5007$\AA\ emission. The region from which we extracted the spectrum is highlighted in the inset panel as the red contours.
The gray shaded area is the error on the cube computed as the quadratic sum of the error extension in the datacube rescaled to match the RMS in the line-free regions of the cube \citep[the error was rescaled by a factor of $\sim$ 3,  see also][]{ Ubler:2023}. 

The G395H spectrum spans the wavelengths between 2.87 -- 5.14 $\mu$m corresponding to 4388 -- 7859~\AA\ rest-frame. In the integrated spectrum, we detect the brightest rest-frame optical emission lines \ha, \hb, \oiii$\lambda\lambda$5007,4959\AA, \nii$\lambda\lambda$6548,6584\AA, \siid\AA\ and \hei\AA.

We fitted the integrated spectrum by modeling it with the sum of a power-law continuum and two Gaussian components for each line, one narrow and one broad. The \siid\ and \hei\ emission lines were fitted only with one narrow line due to the low S/N. We found that one Gaussian profile is enough to reproduce the observed profile of these low S/N lines.
We tied the kinematics of each set of line components (narrow and broad) to have the same velocity and line width. 
For the narrow component, we left the line width free to vary in the range $0<\sigma<250$ \kms, while the broader component was constrained to have a line width larger by at least 20\% than the narrow one, and we imposed the amplitude of the broad component to be lower than 50\% of the narrow one, following \cite{Carniani:2023}. 
The ratios of the intensities of the \oiii\ and \nii\ doublets have been furthermore constrained to atomic physics motivated ratios \citep[see][]{Osterbrock:2006} of \oiii$\lambda$5007/\oiii$\lambda4959 = 2.98$ and \nii$\lambda$6584/\nii$\lambda6548 = 2.94$.

We have also fitted the spectrum by adopting only one Gaussian component for each emission line, but this model has been revealed to be inadequate to reproduce the data. Indeed, the Bayesian information criteria (BIC) test\footnote{By assuming a Gaussian noise, $\rm BIC = \chi^2 + k \ln(N)$ where \textit{k} is the number of free parameters of the fit, and N is the number of data points. } \citep{Liddle:2007} highly favors the two components model with $\Delta \rm BIC = \rm BIC_1 - BIC_2  > 10$ \citep{Kass:1995}, where $\rm BIC_1$ and $\rm BIC_2$ are the values of the BIC for the fit with one and two components for each emission line, respectively.

\begin{table}
\caption{R2700 line fluxes.}             
\label{tab:r2700_fluxes}      
\centering                          
\begin{tabular}{|l|c|}        
\hline\hline   
Line & Flux $\times 10^{-20}[\rm erg$ $\rm s^{-1} cm^{-2}]$ \\    
\hline                        
   \ha -- Narrow & 1617 $\pm$ 35  \\      
   \hb -- Narrow &  421 $\pm$ 18   \\
   \oiii$\lambda$5007\AA\ -- Narrow & 2268 $\pm$ 42\\
    \nii$\lambda$6584\AA\ -- Narrow & 206 $\pm$   17  \\ 
\hline 
   \ha -- Broad & 986 $\pm$ 75 \\      
   \hb -- Broad  &  313 $\pm$ 56   \\
   \oiii$\lambda$5007\AA\ -- Broad  & 1297 $\pm$ 85\\
    \nii$\lambda$6584\AA\ -- Broad  & 297 $\pm$  41   \\ 
\hline                                   
   \hei\AA\   & 97 $\pm$ 32    \\
\sii $\lambda$6717\AA & 122 $\pm$ 9\\
\sii $\lambda$6731\AA & 107 $\pm$ 9\\

\hline
\hline
Component & FWHM [km s$^{-1}$]\\
\hline

   Narrow  & 180 $\pm$ 2    \\
   Broad & 430 $\pm$ 10    \\

\hline
\end{tabular}
\end{table}

In the bottom panels of Fig. \ref{fig:integrated_spectrum} we show the best-fitting results and the residuals of the integrated spectrum analysis for the \hb\ and \oiii\ complex on the left, and the \ha, \nii, and \sii\ complex on the right. The flux of each emission line and the FWHM of the narrow and broad components are reported in Table \ref{tab:r2700_fluxes}.
This analysis allows us to compute the integrated properties of the source and compare it with all previous studies that could not resolve the galaxy morphology.
From the narrow component, we estimated a redshift of $5.54455 \pm 0.00002$ for the galaxy, which is in agreement with the redshift measured from the \cii\ emission \citep{Bethermin:2020, Herrera-Camus:2021}.
On the other hand, a broader component ($ \rm FHWM \sim 430$ \kms) is clearly visible and needed to reproduce all the emission line profiles. 
This component is slightly blueshifted with respect to the narrow component by $v_{\rm broad}-v_{\rm narrow} = - 24 \pm 3$ \kms. 
The FWHM of the broad component is comparable with the \cii\ one, observed by \cite{Herrera-Camus:2021} in this galaxy, but we measured a blueshift, instead of the redshift measured by \cite{Herrera-Camus:2021}. We subsequently discuss the possible interpretation of this component in Sect.  \ref{dsec:discussions}, and we speculate about it being an outflow in Sect \ref{sec:outflow}, where we compute the outflow properties.

By assuming case B recombination, the theoretical ratio between the \ha\ and \hb\ fluxes is $\rm F_{H\alpha}/F_{H\beta}= 2.86$ \citep[for a temperature T=10$^4$ K and a density $n_e = 10^2$ cm$^{-3}$, see][for details]{Osterbrock:2006}. Assuming the \cite{Calzetti:2000} attenuation law ($R_V$ = 4.05), we estimated the dust attenuation of the source from the measured Balmer decrement.
We obtain a value of $A_V =  1.0 \pm 0.2$ for the narrow component nebular dust attenuation. 
By using the same method we also computed the obscuration of the broad component, which is $A_V =  0.3 _{-0.3}^{+0.9}$.

To compute the star formation rate (SFR) of the galaxy, we used the calibration by \cite{Kennicutt:2012} that exploits the \ha\ emission line luminosity. We corrected the narrow \ha\ flux by the dust obscuration reported above, hence we computed an integrated $\rm SFR = 77_{-16}^{+19}$ \msun\ yr$^{-1}$, which is consistent within the error with the one derived from the SED fitting by \cite{Faisst:2020} by exploiting optical and UV photometry, of $41^{+35}_{-15}$ \msun\ yr$^{-1}$.

By exploiting the ratio between the fluxes of \sii$\lambda$6717\AA\ and \sii$\lambda$6731\AA,  we computed the electron density of the source \cite{Sanders:2016} and we obtain a value of  $ n_e = 270_{-200}^{+260}$ cm$^{-3}$.
The estimated electron density is, within the large uncertainties, in agreement with the one obtained from \siid\ from a sample of SF galaxies at $2.7<z <6.3$ \citep{Reddy:2023}, 
and other high-$z$ galaxies targeted in individual NIRSpec-IFS studies \citep{Jones:2024, Lamperti:2023, RodriguezDelPino:2024}. 


\subsection{Spectral fitting of the R100 cube}

\begin{figure}
    \centering
    \includegraphics[width=\hsize]{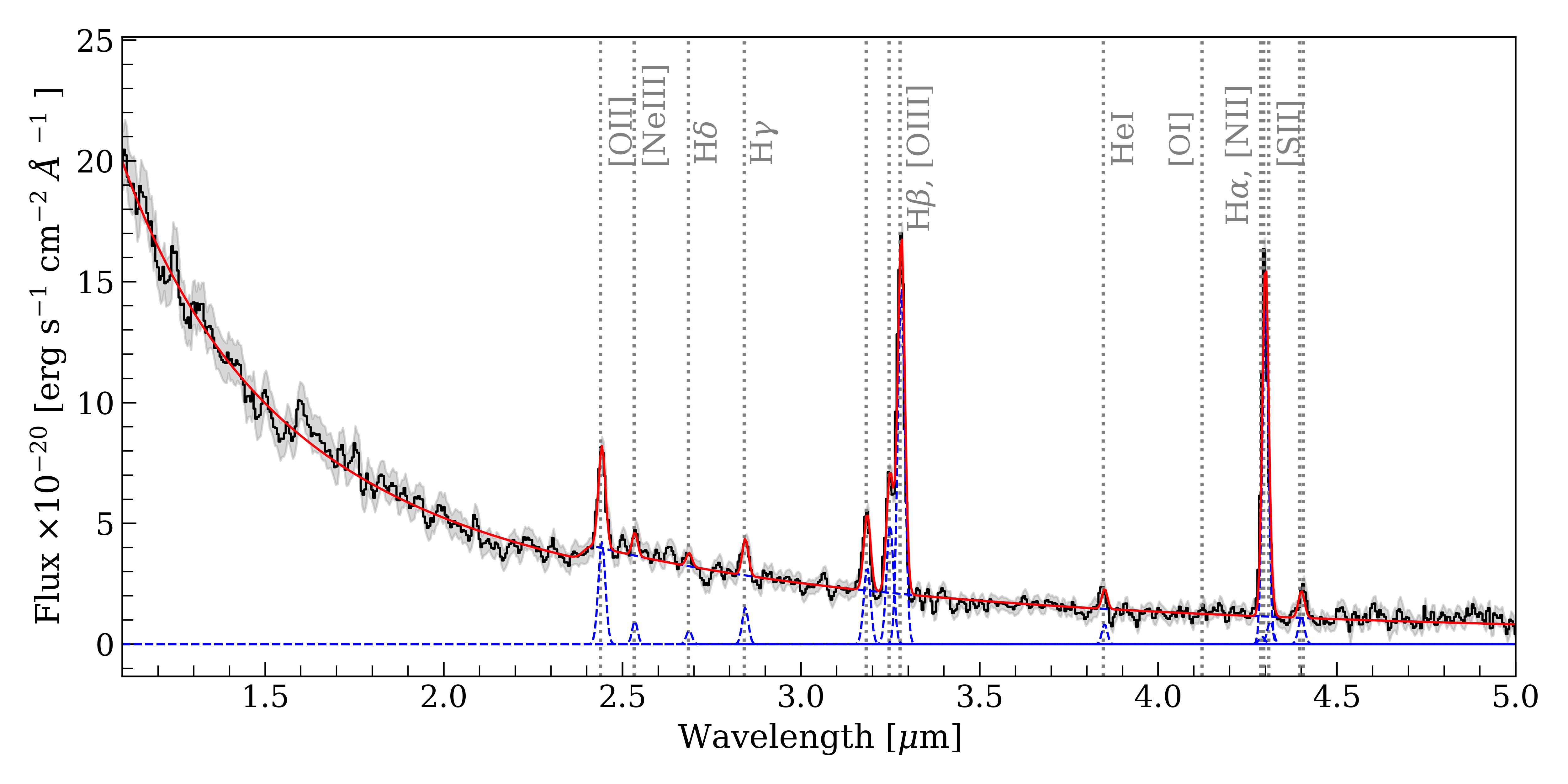}
    \caption{Prism spectrum integrated from the region marked by red contours in Fig. \ref{fig:integrated_spectrum}, and best-fit results.
    In black we show the data, while the gray shaded region represents the 1$\sigma$ error. In blue we show the best-fit emission lines, and in red the best-fit model. }
    \label{fig:fit_prism}
\end{figure}

\label{sec:analysis_R100}

Following the same approach as for the high-resolution cube, we extracted the spectrum from the prism cube from the same aperture as the R2700 one.
We show the R100 integrated spectrum in Fig. \ref{fig:fit_prism}.
The prism spans the wavelength range between 0.6 and 5.2~$\mu$m allowing us to probe both the rest-frame UV and the optical region of the spectrum. In particular, we can now probe the rest-frame wavelengths < 4500 \AA, which are inaccessible in the R2700 cube.
In the new range of wavelengths, we detect the \oiid, \neiii\, and H$\gamma$ emission lines, as well as a blue UV continuum.
We fitted the integrated spectrum as a sum of a continuum emission and a set of emission lines with single Gaussian profiles, since the lower resolution of the spectrum does not allow us to distinguish between the broad and the narrow component, as we did for the R2700 cube.
To account for the Balmer break, we modeled the continuum emission as two different power laws with a discontinuity at $3645 \AA$ rest frame, then we convolved the spectrum with the line spread function of the prism \citep{Jakobsen:2022}. 
The velocities of the lines were tied together, but the velocity dispersion was left free to vary
in order to allow for the varying spectral resolution of the PRISM with wavelength. It is important to note that the lines are unresolved in the prism spectrum, hence the line width only reflects the instrument's resolution.

In addition, we set the flux ratio of the \oiii\ and the \nii\ doublets according to atomic physics prescriptions (see Sec \ref{sec:analysis_R2700}).
 We use only one Gaussian component to fit the \siid\ and the \oiid\ doublets due to the low resolution, which does not allow us to deblend the contribution from each line of the doublet. Therefore, we obtain only the total flux of the doublet.
Similarly, the \nii\ emission lines are blended with the \ha\ emission line, hence it is not possible to disentangle their contribution, and the \ha\ flux is not as reliable as the one estimated from the high-resolution spectrum.

We report the line fluxes measured from the R100 spectrum in Table \ref{tab:r100_fluxes}, and note that within the errors, they agree with those reported for the R2700 cube (after summing the contribution from the narrow and broad components). This demonstrates that the flux calibration largely agrees between the two cubes. 
The only exception is the flux for the \ha\ and \nii\ emission lines, which are blended together, however the summed flux agree within the uncertainties. 
From the expected line ratio of $F_{\rm H\gamma}/F_{\rm H\beta} = 0.466$, we estimated a dust extinction of $A_V = 0.7^{+1.4}_{-0.7}$ assuming that the \oiii$\lambda$4363\AA\ emission line flux, which is blended with the \hg\ in R100 observations, is zero. We find that this value agrees within the large uncertainties with the one computed from the \ha\ and \hb\ ratio from the grating. From now on we will use as fiducial value of $A_V$ the one inferred from the R2700 grating.

\begin{table}
\caption{R100 line fluxes. }             
\label{tab:r100_fluxes}      
\centering                          
\begin{tabular}{|l|c|}        
\hline\hline   
Line & Flux $\times 10^{-20}[\rm erg$ $\rm s^{-1} cm^{-2}]$ \\    
\hline                        
   \ha  &$3032 \pm  110$  \\      
   \hb  &  $761 \pm  36$   \\
    \hg  &  $325 \pm  35$  \\

   \oiii$\lambda$5007\AA\  & $3663 \pm  71$\\
    \nii$\lambda$6584\AA\  & 200 $\pm$   91 \\ 
    \hei\AA\  & $158 \pm 29$\\
    \neiii\AA\ & $185 \pm 40$ \\
    \oiid \AA\ &   $1153 \pm  90$  \\
    \siid \AA\ &   $242 \pm  35$  \\ 

\hline
\end{tabular}
\end{table}

By exploiting the strong-line metallicity diagnostics based on rest-frame optical line ratios, we can calculate the gas phase metallicity of the galaxy.
We adopted the calibrations from \citet{Curti:2017, Curti:2020}, recently revised for high-$z$, low-metallicity galaxies \citep{Curti:2024, Laseter:2024}.
In particular, we exploited the diagnostics R3 = \oiii$\lambda$5007\AA/\hb, R2 = \oiid/\hb, $\rm \hat{R}$ = $\rm 0.47 \,R2+0.88 \,R3$, O32 = \oiii/\oiid, N2 = \nii/\ha, and O3N2 = (\oiii/\hb)/(\nii/\ha) to estimate the abundance of oxygen $\rm \log(O/H)$.

The \oiid\ emission line doublet is only covered by the R100 observations, which do not allow us to disentangle the broad component contribution. Hence, we can estimate a global metallicity by assuming the total line flux for each line (from the PRISM for \oiid, narrow+broad from the grating for the other lines). We corrected each line flux for the dust obscuration computed from the Balmer decrement as the ratio between the total fluxes (narrow+broad) of  \ha\ and \hb\ in the grating ($A_V = 0.7 \pm 0.3$).
We estimate a value of $\rm 12 + \log(O/H) = 8.34 \pm 0.08$ for the entire galaxy. By using only the R3, N2, and O3N2 calibrators (i.e., those provided by the grating spectrum) we are also able to estimate the galaxy versus broad metallicity. We obtain an abundance of
$\rm 12 + \log(O/H)_{\rm narrow} = 8.3 \pm 0.1$ for the entire galaxy and $\rm 12 + \log(O/H)_{\rm broad} = 8.4 \pm 0.1$ for the broad component. 
 Both values are consistent with the value we infer by adding the calibrators that exploit the \oiid\ lines.
 
The value that we obtain is in agreement with metallicity obtained for other high-$z$ galaxies with similar stellar mass ($\log(M_\star/M_\odot) = 9.75^{+0.05}_{-0.10}$ for HZ4, see Sect. \ref{sec:SED}) and merging systems \citep{Nakajima:2023, Curti:2024, Sanders:2024, Venturi:2024}. 
Assuming a solar oxygen abundance of $\rm 12 + \log(O/H)_{\rm narrow} = 8.69$ \citep{Asplund:2021},
 we obtain a metallicity of $Z_{\rm narrow} = 0.4\pm 0.1~Z_\odot$ and $Z_{\rm broad} = 0.5\pm 0.1~Z_\odot$ for the narrow and broad components, respectively.

\section{Spatially resolved analysis}

\label{sec:analysis_resolved}

\subsection{Channel maps and continuum emission}

\begin{figure*}
    \centering
    \includegraphics[width=\hsize]{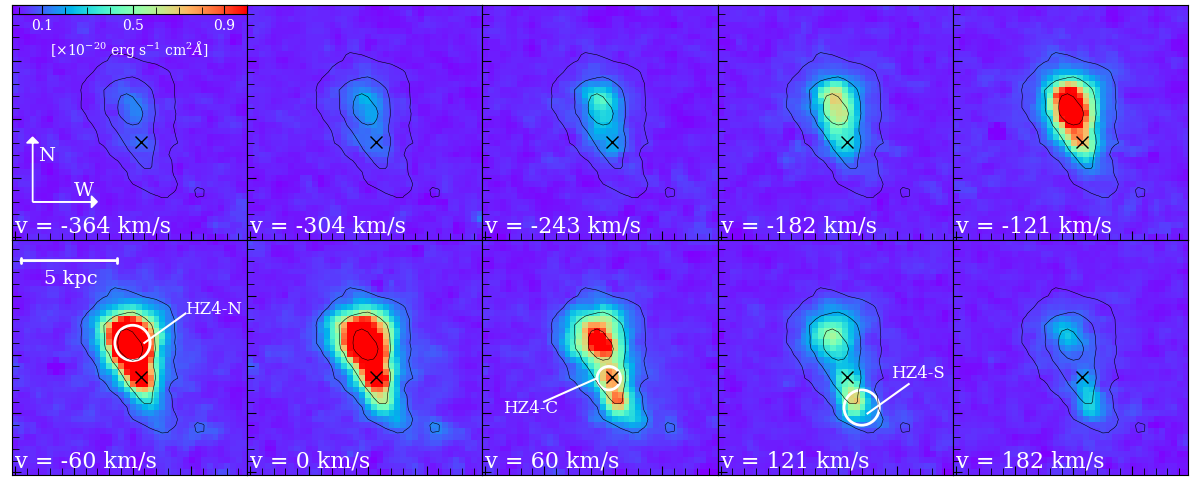}
    \caption{Channel maps from the R2700 cube targeting the \oiii$\lambda$5007\AA\ emission line. 
    Black lines are the contour of the 5, 25, and 75 times the RMS level computed in the line-free regions of the cube for the emission at $v=0$.  The velocity marks the central velocity of the channel.  The black cross shows the location of the \cii\ emission peak (see Fig. \ref{fig:astrometry}). The white circles represent the position of the components of HZ4.}
    \label{fig:channel_maps}
\end{figure*}

The high-resolution NIRSpec R2700 datacube shows that the galaxy has different components at different velocities.
This can be seen from the ten channel maps of \oiii\ at different velocities shown in Fig.~\ref{fig:channel_maps}.
We set the zero velocity at the systemic redshift measured from the narrow component of the integrated R2700 spectrum.
In each channel map, we show the contours at 5, 25, and 75$\sigma$ of the map at $v=0$ \kms\ to highlight the different morphologies of the \oiii\ emission at different velocities.

The $v=0$ map shows an elongated morphology, with a central peak that coincides with the peak shown in the UV observations (see also Fig. \ref{fig:astrometry}) and an extended tail in the southern direction.
From the map at $v = 60$ \kms\ it is evident that the elongation of the source is due to the presence of three components.
In particular, a large southern clump is visible in the redshifted channels ($v>60$ \kms). We label this component as HZ4-South. 
The presence of these different components hints that the galaxy may be a galaxy merger or a clumpy galaxy, as observed for other high-$z$ sources \citep{Carniani:2018, Arribas:2023, Jones:2023, Venturi:2024}.
On the other hand, in the blueshifted channels, a component extending in the north direction and peaking in the maps at $v<-182$\kms\ slightly northern than the 75$\sigma$ contours peak at $v=0$ becomes evident.
This component hints at the broad component that we will investigate in the next Sections.

\begin{figure*}
    \centering
    \includegraphics[width=\hsize]{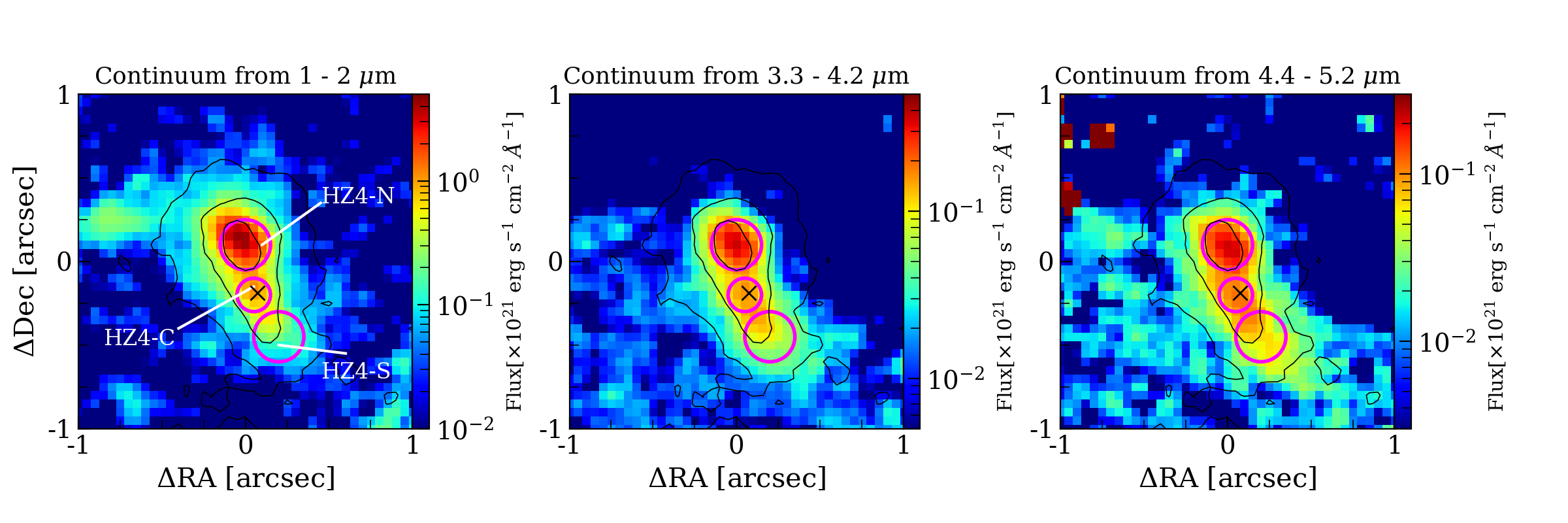}
    \caption{Continuum emission from the R100 JWST/NIRSpec datacube.
    From left to right the average continuum emission between 1--2 $\mu$m (1530--3058 \AA rest-frame), 3.3--4.2 $\mu$m (5045--6422 \AA rest-frame), 4.4--5.2 $\mu$m (6730--7950 \AA rest-frame), respectively.
    The magenta circles represent the apertures from which we extract the spectrum for the north, central, and southern components for the analysis.
    The black contours represent the \oiii\ emission at 5, 25, and 75 $\sigma$ (see Fig. \ref{fig:astrometry}).
    The black cross shows the location of the \cii\ emission peak (see Fig. \ref{fig:astrometry}).
    The $x$ and $y$ axes represent the displacement in arcsec with respect to the galaxy center (RA = 09h58m28.5s, Dec= +02d03m06.7s, see Sect. \ref{sec:introduction}).
}
    \label{fig:continuum}
\end{figure*}

We also looked at the continuum emission from these sources to identify these three different components.
Fig. \ref{fig:continuum} shows the continuum emission in three different wavelength ranges obtained from the NIRSpec R100 cube.
In the left panel, we show the emission of the rest-frame UV continuum (1--2 $\mu$m observed wavelength or 1530–3058 \AA  rest-frame). The UV emission morphology is similar to the one of the \oiii\ line (shown as the black contours); the location of the two peaks coincides.
Another source appears visible leftward of the target. We identify this source as a foreground source at $z\sim 2.6$ and we discuss in more detail its properties in Appendix \ref{appendix:foregroundsource}.
From the rest-frame optical continuum in the observed range 3.3--4.2 $\mu$m (5045–6422 \AA rest-frame) and 4.4--5.2 $\mu$m (6730–7950 \AA rest-frame), shown in the central and right panels of Fig. \ref{fig:continuum}, respectively. Here the southern, redshifted clump, already visible in the channel maps at $v = 121$~\kms, is now evident.
This clump is very faint in \oiii\ emission (and any other emission line) but appears visible in continuum.
The location of the central, small clump already identified in the channel maps at $v = 60$~\kms\ is coincident with the location of the peak of the \cii\ emission (see Fig. \ref{fig:astrometry}) and the peak of the SFR discussed in Sect. \ref{sec:analysis_resolved}.

From the maps presented in this section, we can distinguish three different galaxy components. 
A northern component (HZ4-North) which is very bright in rest-frame UV and optical continuum and \oiii\ emission, a central component (HZ4-Central) which is bright in \cii\ emission, and a redshifted southern component (HZ4-South), which is faint in \oiii\ emission but with an extended tail visible only in the rest-frame optical continuum.

\subsection{Spaxel-by-spaxel R2700 results }

\begin{figure*}
    \centering
    \includegraphics[width=\hsize]{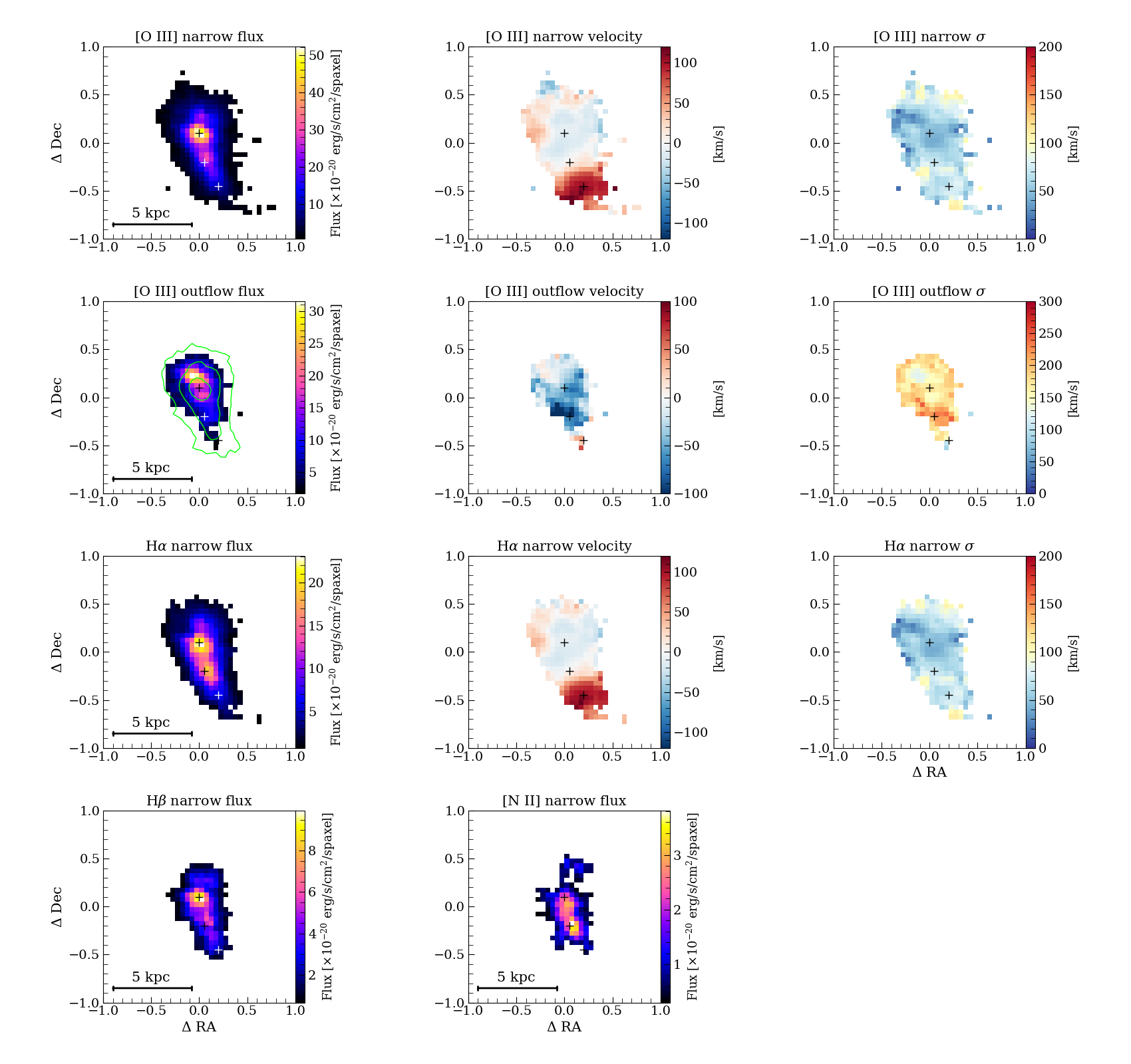}
    \caption{Results for the  spaxel-by-spaxel fitting of the R2700 cube.
    In the first row, we show from left to right the \oiii$\lambda$5007\AA\ flux, velocity, and velocity dispersion maps of the narrow component, respectively.
    In the second row, we show the same but for the \oiii\ broad component. In the \oiii\ broad flux map, we show the 5, 25, and 75$\sigma$ contours of the narrow \oiii\ flux in green.
    In the third row we show from left to right the \ha\ flux, velocity, and velocity dispersion maps of the narrow component, respectively.
    In the bottom panels, we show from left to right the flux maps of the \hb, and \nii$\lambda$6584\AA\ narrow components, respectively.
    The plus signs show the location of the three components identified in Figs. \ref{fig:channel_maps} and \ref{fig:continuum}.}
        \label{fig:moment_maps}
\end{figure*}

The high S/N of the cube at the spaxel level allows us to perform a spatially resolved analysis of the galaxy. We fitted the cube at a spaxel-by-spaxel level with both a single Gaussian component for each emission line and two Gaussian components, following the same approach discussed in Sect. \ref{sec:analysis_R2700}. 
We then selected for each spaxel the best fit by performing the BIC test and selecting the model with two components for each line only when $\rm BIC_1 - BIC_2 > 10$, which represents positive evidence toward the model with two components \citep{Kass:1995}. In the other cases, we selected the model with only one component, as the addition of the broader component was not significantly improving the quality of the fit. We excluded all the spaxels with $S/N < 3$ for each emission line. The S/N is computed as the ratio between the flux of the best-fit emission line and the resulting error from the fit.

The results of the fitting (flux, velocity, and velocity dispersion maps) for the narrow and broad components are shown in Fig. \ref{fig:moment_maps}.
The galaxy morphology traced by the rest-frame optical emission lines differs from what is observed in the rest-frame UV continuum and the \cii\ FIR emission line, as already seen in the channel maps in Fig. \ref{fig:channel_maps}.
The narrow \oiii\ flux map shows a bright clump in the northern region (HZ4-North). This clump is in the same location as the bright emission observed in the F160W filter from HST (see also Fig. \ref{fig:astrometry}), which is tracing rest-frame UV emission. Additionally, the \oiii\ map shows an extended tail, in the southwest direction. The tail in \oiii\ shows the presence of a second faint component (HZ4-Central), which is instead more prominent and visible in the \ha\ map, and is also where the \cii\ peak.
This suggests that the second component is likely tracing a region with higher dust obscuration, as we discuss later in this section.

The narrow component velocity field (top-central panel of Fig. \ref{fig:moment_maps}) shows a different picture from what was seen in ALMA observations. Whereas the \cii\ kinematics is consistent with a rotating disk \citep{Herrera-Camus:2021}, the ionized gas shows a more complex kinematic pattern, also due to the higher spatial resolution of the JWST data. In the northern region of the galaxy, we observe a relatively shallow velocity gradient ranging from 50 to --20 \kms, while the southern region is redshifted by $\sim 100-130$ \kms. The clumpy morphology of the galaxy, combined with the observed abrupt velocity gradient in the southern part, suggests that the galaxy is undergoing a merger event rather than being a rotating disk as inferred from ALMA observations \citep{Jones:2021, Herrera-Camus:2022, Parlanti:2023}.

In the second row of Fig. \ref{fig:moment_maps} we show the broad component maps. In the broad flux map, we also show as green lines the narrow component 5, 25, and 75$\sigma$ contours.
The morphology of the broad emission shows an elongated shape in the same direction as the narrow component emission. 
The brighter region is also in the northern part, but the peaks of the broad and the narrow emission are not in the same location. It is evident that the narrow component peaks $\sim 0.1$\arcsec\ to the south of the broad emission, which corresponds to a projected distance of 0.6 kpc.
The velocity map shows that in the majority of the spaxels the broad component is blueshifted by up to --130 \kms\ relative to the galaxy-integrated systemic redshift computed from the narrow component. The velocity shift relative to the narrow component increases in the southern part of the broad component. The southern part coincidentally has also the highest velocity dispersion.

In Fig. \ref{fig:out_velocity_sfr},  we assumed that the broad component is tracing an outflow, and we show the de-projected outflow velocity calculated as $v_{\rm out} = |\Delta v_{\rm narrow, broad}| + 2\sigma_{\rm out}$. The velocity reaches its maximum of approximately 600 \kms\ at the position of the HZ4-Central component. This location coincides with the highest $\Sigma_{\rm SFR}$, which is indicated by dashed contours in Fig.~\ref{fig:out_velocity_sfr} and will be discussed at the end of this section.
The presence of the higher velocity at the location of the maximum SFR and the fact that the peak flux of the broad component does not coincide with the peak of the narrow one lead us to interpret that this broad component may be an outflow that is launched from the HZ4-Central component. Another possibility could be that we are observing multiple winds driven by the SF regions distributed in the system or, we are witnessing merger flows or shocks. 
We will discuss the possible scenarios in Sect.~\ref{dsec:discussions}, while we will explore more in detail the outflow possibility in Sect.~\ref{sec:outflow}.

\begin{figure*}
    \centering
    \includegraphics[width=\hsize]{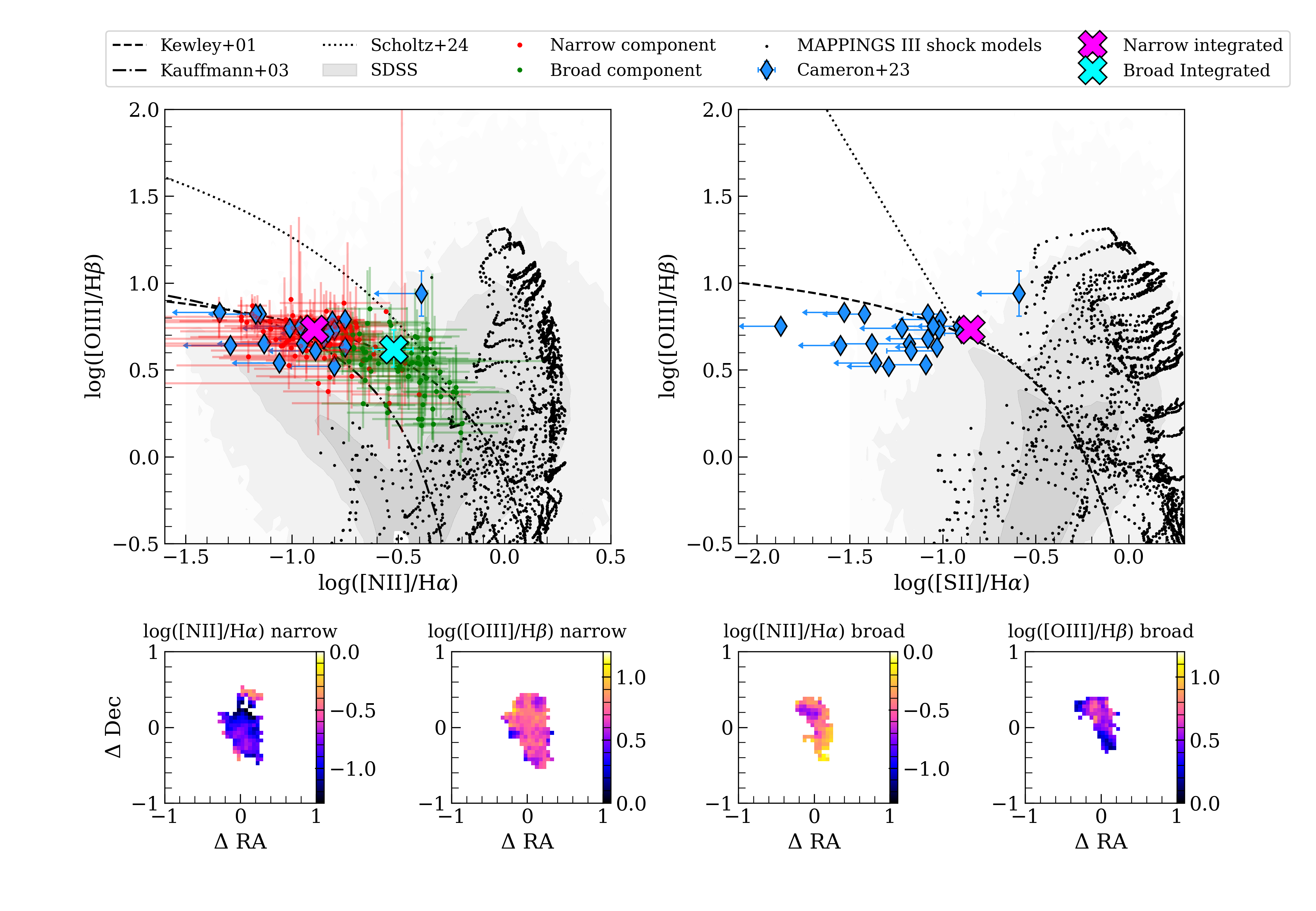}
    \caption{
    Top panels: BPT diagram on the left and  VO87 diagram on the right. 
    As pink and turquoise crosses we show the location on the BPT diagram of the narrow and the broad component for the integrated spectrum. 
    As red and green points we report the location in the BPT diagram of each spaxel for the narrow and broad component, respectively. 
    The gray shaded region shows the location of SDSS galaxies, while blue diamonds show the location in this diagram of low-metallicity high-redshift ($z\sim5.5-9.5$) star-forming galaxies from \cite{Cameron:2023}. 
    Black dots show the location of the MAPPINGS III shock model from \cite{Allen:2008} with shock velocities between 200 and 1000 km/s, densities in the range $0.01~ {\rm cm}^{-3} \leq n \leq 1000~{\rm cm}^{-3}$, solar abundances, and magnetic field between $10^{-4}$ and $10$ $\mu$G.
    As black dashed and dash-dotted lines, we report the demarcation line for the local star-forming galaxies and AGN samples by \cite{Kewlwy:2001} and \cite{Kauffmann:2003}. 
    As the dash-dotted black line we show the demarcation line between high-z low-metallicity galaxies and AGNs by \cite{Scholtz:2023}. AGNs are right-ward of the line, while on the left there is a mixed population of AGNs and SF galaxies.
    Bottom panels: from left to right the logarithm of N2, R3 in each spaxel for the narrow component, and the logarithm of N2, R3 in each spaxel for the broad component.}
    \label{fig:BPT}
\end{figure*}

\begin{figure*}
    \centering
    \includegraphics[width=\hsize]{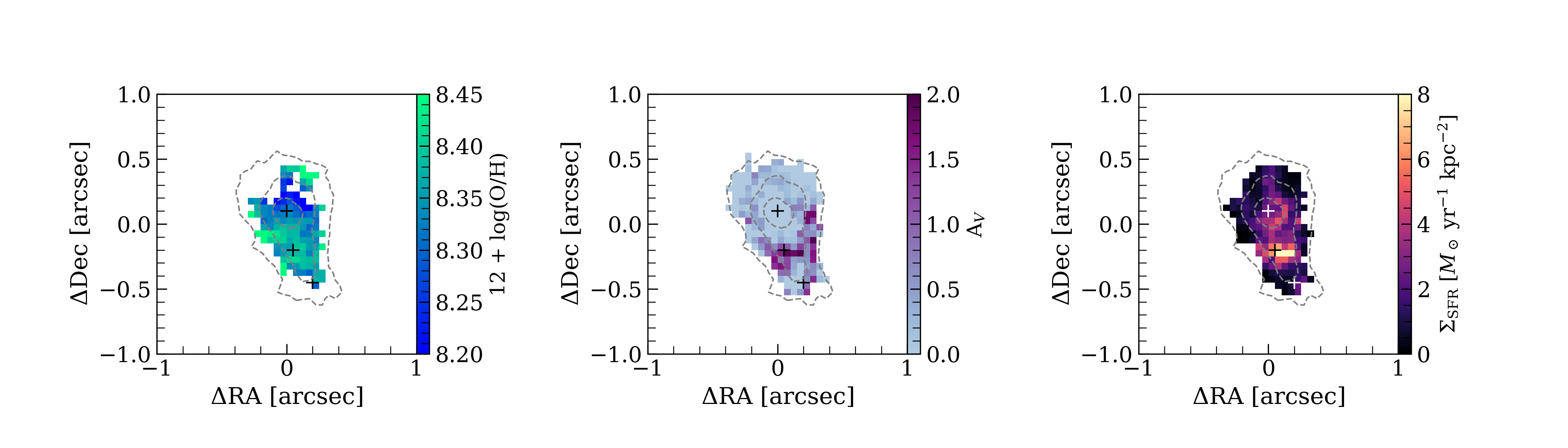}
    \caption{Spatially resolved properties of the narrow component.
    From left to right: gas-phase metallicity ($\rm 12 + \log(O/H)$), dust extinction ($A_V$), and dust-corrected star formation rate surface density ($\Sigma_{\rm SFR}$) maps. The plus signs show the location of the three components identified in Figs. \ref{fig:channel_maps} and \ref{fig:continuum}. The gray contours represent the 5, 25, and 75$\sigma$ of the narrow \oiii\ flux from Fig. \ref{fig:moment_maps}.}
    \label{fig:dust_sfr}
\end{figure*}

In Fig. \ref{fig:moment_maps} we also present the \nii\ flux map, which shows a different morphology from the flux maps of the other emission lines. We can highlight the presence of two peaks in the \nii\ emission, one associated with the northern component, and one associated with the central one. Differently from the \ha\ and \oiii\ the brightest \nii\ emission arises from the central region.
These differences can be a sign of a different metallicity between the two galaxies or a different ionization mechanism, with the presence of a harder radiation field arising in the center.

To study this latter possibility we exploit the Baldwin-Phillips-Terlevich (BPT; \citealt{Baldwin:1981}) and the VO87 \citep{Veilleux:1987} diagnostic diagrams, which are sensitive to the ionizing source in a galaxy, between star formation in HII regions and AGN photons.
In particular, the first one is based on the line ratios of \nii/\ha\ and  \oiii/\hb\, while the latter uses the  \siid/\ha\ and \oiii/\hb.
In the top-left panel of Fig. \ref{fig:BPT} we report the spatially integrated measurement of the galaxy in the BPT diagram for the narrow and the broad component, together with the location of each spaxel.
We observe that the integrated emission and the majority of the spaxels, of the narrow component, lie on the \cite{Kewlwy:2001} demarcation line, together with the low-metallicity $z\sim$5.5--9.5 galaxies from \cite{Cameron:2023}.
On the other hand, the majority of the broad component points lie in the composite region of the BPT between the demarcation lines of \citet{Kewlwy:2001} and \citet{Kauffmann:2003} that are based on the SDSS galaxy sample (gray shaded region).
Low-redshift observations have shown that in the presence of shocks induced by starburst-driven winds, or by tidal flows due to a merger, the line ratios can shift toward the composite region even in the absence of an AGNs \citep{MonrealIbero:2010,Rich:2011, Rich:2015, Medling:2018}. 

To check for the presence of shock, we over plot the MAPPINGS III shock-ionization models \citep{Allen:2008} with densities of $n=100$ and 1000  cm$^{-3}$, velocities ranging from 200 to 1000 km/s and magnetic fields from 10$^{-4}$ to 10 $\mu$G. We observe that the shock models cannot reproduce the location of the broad line ratios in the BPT diagram, with the exception of just a few extreme points.
We note that the presence of a large-scale galactic outflow can shock-excite the galaxy interstellar medium \citep{Ho:2014}, increasing the \nii/H$\alpha$ line ratio.

The demarcation lines used in the BPT diagram for low-$z$ sources have proved to be unreliable at high-$z$ ($z>3$).
Due to the increase of the ionization parameter and the decrease of the metallicity at high redshift, both galaxies and AGNs may end up in the same location in the BPT diagram \citep{Nakajima:2022, Ubler:2023, Harikane:2023, Maiolino:2024, Scholtz:2023}.
We then used the new demarcation line by \citet{Scholtz:2023}, shown as a dash-dotted line in the plot. 
According to this new line, there are no points that lie in the AGNs region, but all points lie in the region in which SF galaxies and AGNs may have the same line ratios.

In the bottom-right panel, we show where the narrow component lies in the VO87 diagram, together with the sample of high-$z$ galaxies from \cite{Cameron:2023} and the local galaxies from SDSS.
The point lies on the demarcation lines by \citet{Kewlwy:2001} and \citet{Scholtz:2023}, therefore we cannot constrain the ionization source. 

Unfortunately, for the wavelength range and spectral resolution available, there are no other diagnostic diagrams that have been proven to be reliable at high redshift \citep{Scholtz:2023, Mazzolari:2024}. 
The absence of He{\sc{ii}}  and other high-ionization lines rule out the presence of a strong and energetic AGN and favor a SF interpretation, although we cannot completely rule out a contribution to ionization from a weak, subdominant AGN.

We display the maps of the \lognii\ and \logoii\ ratios for the narrow and broad components in the top row of Fig. \ref{fig:BPT}. For the narrow component, a gradient in the values of \lognii\ is evident, with the ratio increasing from north to south, while the values of \logoii\ are almost constant. The broad component, on the other hand, shows a gradient also in the \logoii\ values, varying from 0.2 to 0.7, while the \lognii\ values range from --0.8 to --0.2.
Since the AGN scenario is disfavored, the observed gradients of the line ratio can be interpreted as a metallicity gradient between the southern and the northern clump with the northern clump being poorer in metals compared to the southern one, given that the \nii/\ha\ ratio decreases as the metallicity decreases \citep{Pettini:2004, Curti:2017, Curti:2020}. Metallicity gradients have been observed at high redshift, especially in clumpy or merging systems \citep{Arribas:2023, RodriguezDelPino:2024, Venturi:2024}. The study of the metallicity gradients for this galaxy will be presented in Curti et al. (in prep.), but here we show the metallicity maps for the narrow component in the left panel of Fig. \ref{fig:dust_sfr}.
We exploit the N2, R3, and O3N2 line ratios and we adopt the calibrations from \citet{Curti:2017} to derive the O/H abundances.
The estimated abundances vary between $\rm 8.2 < 12 + \log(O/H) < 8.5$. The lowest metallicity values are near the center of the northern and central components, while for the southern component, the lack of detection of \nii\ emission does not allow us to constrain the metallicity.

Having excluded the presence of an AGN, we can also estimate the SFR surface density ($\Sigma_{\rm SFR}$) at a spaxel level by exploiting the \ha\ dust-corrected luminosity.
First, we estimated the dust attenuation at a spaxel level thanks to our detection of \ha\ and \hb\ emission lines.
By assuming case B recombination (T$_e$=10,000~K and $n_e=100$ cm$^3$), the theoretical ratio between the \ha\ and \hb\ flux is $\rm F_{H\alpha}/F_{H\beta}= 2.86$ \citep{Osterbrock:2006}. We measured a spaxel-by-spaxel ratio ranging from 2.86 to 5.60. Assuming the \cite{Calzetti:2000} attenuation law, we estimate a dust attenuation of $0\leq A_V\leq 2.5$. In the central panel of Fig. \ref{fig:dust_sfr} we show the map of dust attenuation across the galaxy.

Interestingly, the northern region, where the line emission and the UV continuum peak, shows little to no dust extinction, similar to what is observed in the southern component. However, there is a zone of high extinction between these two regions where the dust attenuation is reaching values of $A_V \sim 2.5$ which is cospatial with the central component.

By correcting the \ha\ flux for the dust extinction and using the calibration from \cite{Kennicutt:2012} to get the SFR from \ha\ luminosity,  we computed the resolved SFR surface density map, which is shown in the right panel of Fig. \ref{fig:dust_sfr}. This shows that there is a region of intense and obscured ongoing star formation in the galaxy, which is coincident with the location of the central component.
By comparing with the \cii\ flux contours shown in Fig. \ref{fig:astrometry}, it can be seen that this region is coincident with the peak of the \cii\ emission, which indeed is a tracer of star formation unaffected by the presence of dust \citep{DeLooze:2011, Pineda:2014, HerreraCamus:2015}.

\section{SED fitting}
\begin{figure*}
    \centering
    \includegraphics[width=\hsize]{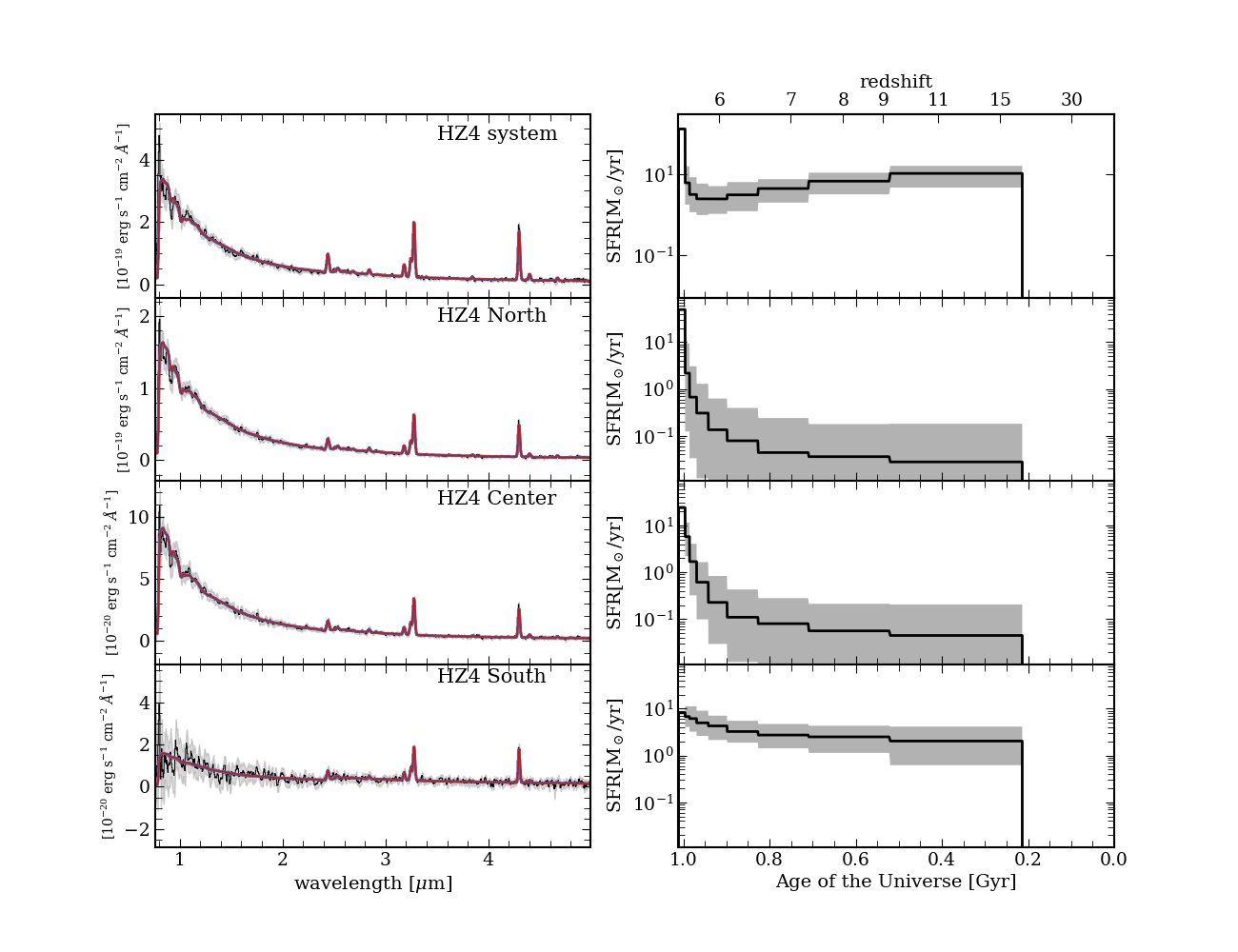}
    \caption{On the left: R100 spectra in black and SED best-fitting result in red extracted from the entire system, the northern clump, the central clump, and the southern clump, from top to bottom, respectively.
    On the right: best-fit star formation history.}
    \label{fig:bagpipes}
\end{figure*}

\label{sec:SED}
The prism data allow us to perform a spectral energy distribution (SED) fitting to study the galaxy's stellar population and infer key galaxy properties such as the stellar mass and the star formation history. 
We used the SED fitting code \verb|Bagpipes| \citep{Carnall:2018}. We adopted the stellar population SED model from \cite{Bruzual:2003}, and we used \verb|Cloudy| \citep{Cloudy:2023} to obtain the models for the nebular emission with a range in the ionization parameter priors of $-3<\log U<0$. 
We assumed a \cite{Calzetti:2000} dust attenuation law, with $A_V$ free to vary between 0 and 4.
We assumed the same extinction for the nebular and stellar components.
We left the metallicity free to vary within the range $0.1 - 1$ $Z_\odot$.
For the star formation history (SFH), we adopt a so-called nonparametric SFH, with continuity priors \citep{Leja:2019} with 10 bins logarithmically spaced from 1 Gyr ($z=5.5$, the redshift of the source) up to 200 Myr after the Big Bang.

First, we performed the SED fitting on the integrated R100 spectrum of the regions in which the S/N on the \oiii\ emission line is larger than 3 (see red region in the inset panel in Fig. \ref{fig:integrated_spectrum}).
We show the spectrum and the best-fit model in the top-left panel of Fig.~\ref{fig:bagpipes}, while we show the SFH in the top-right panel. The properties resulting from the SED fitting are reported in Table \ref{tab:integratedproperties_bagpipes}.
We infer a total stellar mass of the system of $\log(M_\star/M_\odot) = 9.75_{-0.10}^{+0.05}$ which is smaller than the previous estimate from the literature of $\log(M_\star/M_\odot) = 10.15\pm 0.15$  of \citet{Faisst:2020}, but in agreement with the one from \citet{Capak:2015} of $\log(M_\star/M_\odot) = 9.67\pm 0.21$.

The SFH for the whole HZ4 system shows an almost constant SFR of a few \sfr\ up to the last 20 Myr, where we see a rapid increase reaching $\sim$ 100 \sfr\ that could be triggered by the merger \citep{Pearson:2019}.
Indeed, the integrated spectrum shows the Balmer break at $\sim 2~ \mu$m, which is indicative of an older stellar population, while the luminous \ha\ is indicative of SFR in the last 10 Myr.

If we extract the spectra for the northern, central, and southern components (see apertures in Fig. \ref{fig:continuum}) we see that the older population is dominant in the southern component, with a well-defined Balmer break, faint continuum UV emission, and faint emission lines. On the other hand, the spectra extracted in the central and northern regions show a bright UV continuum.
This difference in the spectra indicates a different stellar population that reflects different formation histories of the northern, central, and southern components (see Fig. \ref{fig:bagpipes}, right panels). 
The southern component started to form stars only $\sim 200$ Myr after the Big Bang and has continued to grow with a slowly increasing SFR, reaching values of SFR = $ 8 \pm  2$ \sfr\ at the time of the observation.
From the SED fitting we obtain a mass-weighted age of the southern galaxy of $322^{+88}_{-108}$ Myr.
On the other hand, the northern galaxy shows a much quicker rise in the SFR, having assembled most of its mass in the last 30 Myr.
The central component has a rising SFH very similar to the one of the northern one. Both the northern and the central components have a very young stellar population, with a mass-weighted age of $<100$ Myr.

The differences in the SFH and the stellar populations support our interpretation that this galaxy is formed by merging systems, rather than three star-forming clumps. 
Cosmological simulations from \cite{Nakazato:2024} show that a merger event can induce the formation of clumps which lead to a bursty star formation with timescales of 10--30 Myr, consistent with what we observe in the northern and central clumps. 

\begin{table}
\caption{Galaxy properties inferred from the SED fitting.}             
\label{tab:integratedproperties_bagpipes}      
\centering                          
\begin{tabular}{|l|c|c|c|}        
\hline\hline   
 & $\log(M_\star/M_\odot)$ & $A_V$ & $\tau$/Myr  \\    
 & (1) & (2) & (3)  \\   

\hline                        
HZ4 System  &$9.75 _{-0.10}^{+0.05}$ & $0.39 \pm  0.04$ & $357^{+69}_{-95}$ \\
 HZ4-North & $8.90 \pm 0.05$ & 0.28 $\pm$ 0.03           &  $27_{-17}^{+43} $\\
HZ4-Center & $8.69 \pm 0.06$ &$0.25 \pm 0.04$           &  $50_{-32}^{+70} $ \\
 HZ4-South&  $9.24 \pm 0.08$&  $0.71 \pm 0.13$          & $322^{+88}_{-108}$\\

\hline
\end{tabular}
\tablefoot{(1) Stellar mass; (2) Dust attenuation; (3) Mass-weighted stellar age.}

\end{table}

\begin{figure*}
    \centering
    \includegraphics[width=\hsize]{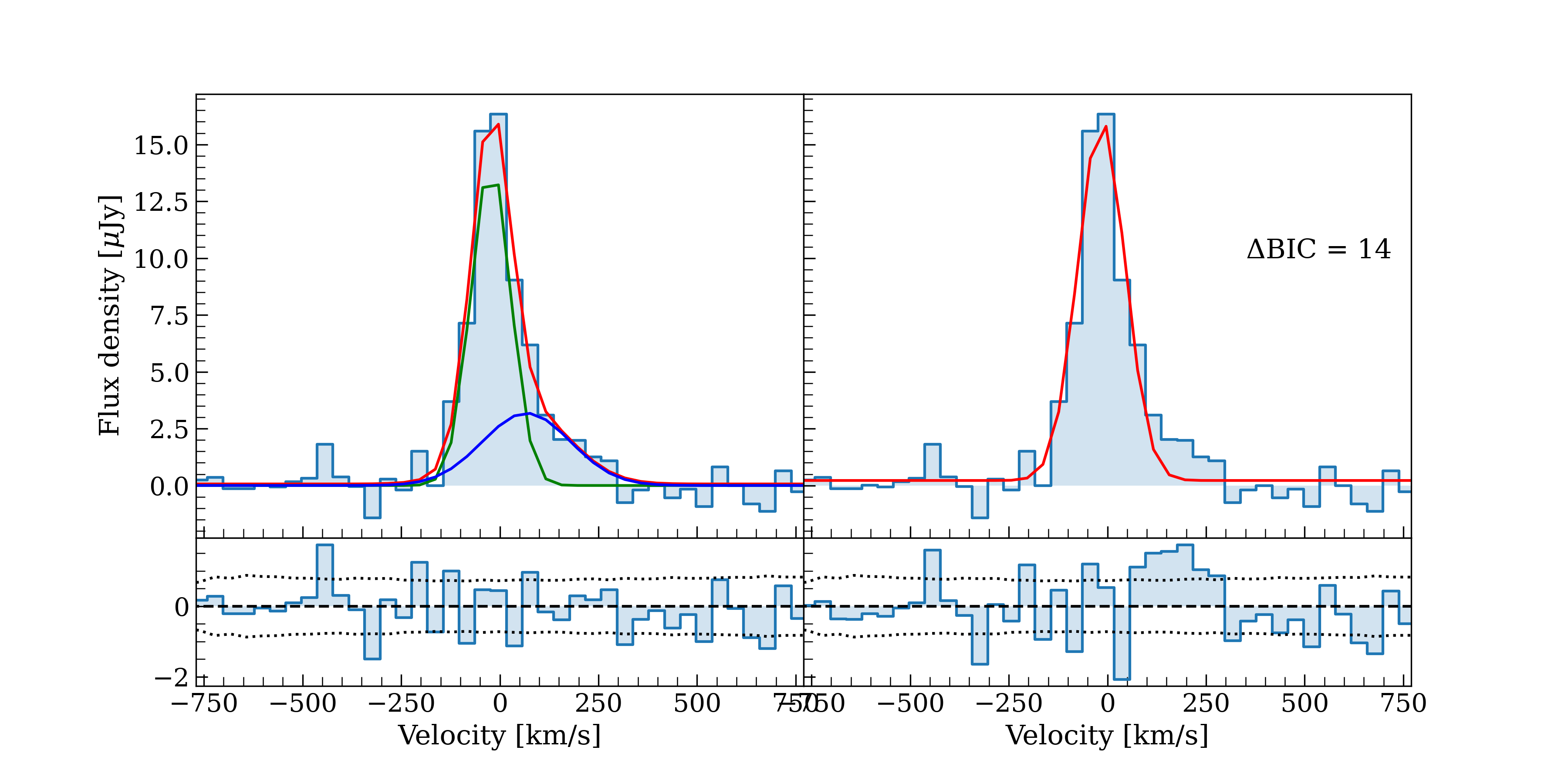}
    \caption{Example of best-fit results for the fitting of the \cii\ emission line in one spaxel. On the left, we show the best-fit results by using two components, on the right the best-fit results by using one component. In the bottom panels, we show the residuals of the fitting procedure. The dotted lines show the RMS computed in the line-free regions of the cube. 
    }
    \label{fig:cii_one_spaxel_fit}
\end{figure*}

\section{ALMA analysis}

\label{sec:ciianalysis}
In this section, we describe the analysis of the ALMA observations of the \cii\ emission line and the dust continuum. 
The ALMA \cii\ cube was analyzed with the same method as the NIRSpec R2700 cube, to detect and possibly spatially resolve the broad component seen in the neutral gas phase. Since the properties of the outflows integrated in one beam are already discussed in \cite{Herrera-Camus:2021}, here we focus only on the spaxel-by-spaxel analysis.
To study the outflow component we exploited the datacube which was reduced with natural weighting, which has a higher S/N at the expense of angular resolution with respect to the Briggs datacube. 
We fitted each spaxel having a S/N of the \cii\ emission greater than 3 with both a single and two Gaussian lines and a constant continuum. The continuum emission was already removed from the cube during the data reduction, but we added this component in the fitting to furthermore remove any residual continuum emission that was not accounted for.
Using a similar method as described for the JWST datacube, for each spaxel we selected the 2-component model only when $\rm BIC_1 - BIC_2 > 5$ and the S/N of both the narrow and the broad component were greater than 3, otherwise the 1-component model was adopted. We note that in this case, we use a slightly lower threshold for the BIC selection due to the lower S/N of the line, but $\Delta \rm BIC >5$ still gives positive evidence in favor of the two Gaussian model.
We applied the same procedure to the Briggs (robust = 0.5) datacube. Unfortunately, the lower S/N of this latter datacube does not allow us to select the outflow component in any spaxel due to its S/N falling below our threshold of 3, even if the $\Delta$BIC was larger than 5 in some spaxels.
But a broad component is still detected integrating into circular apertures as also shown by \citet{Herrera-Camus:2021}.

\subsection{\cii\ spaxel-by-spaxel results}

\begin{figure*}
    \centering

    \includegraphics[width=\hsize]{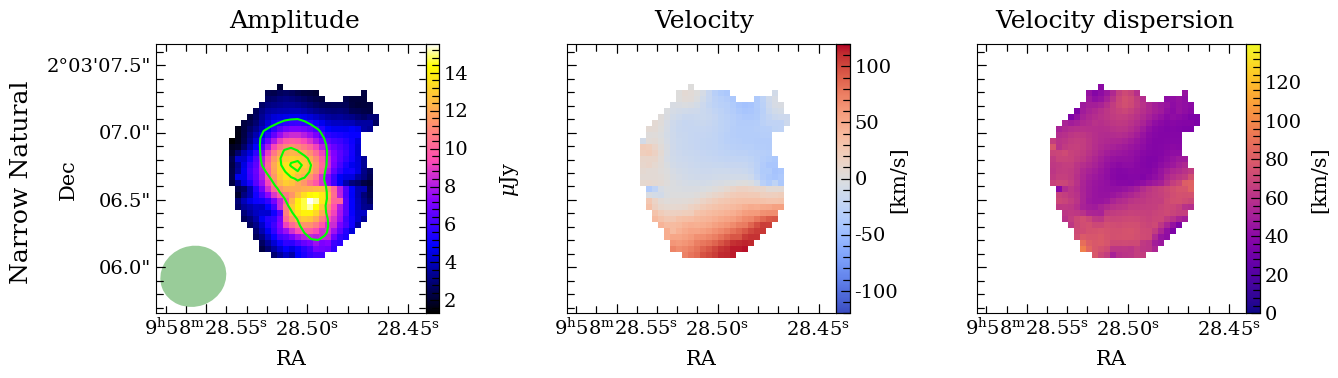}
    \includegraphics[width=\hsize]{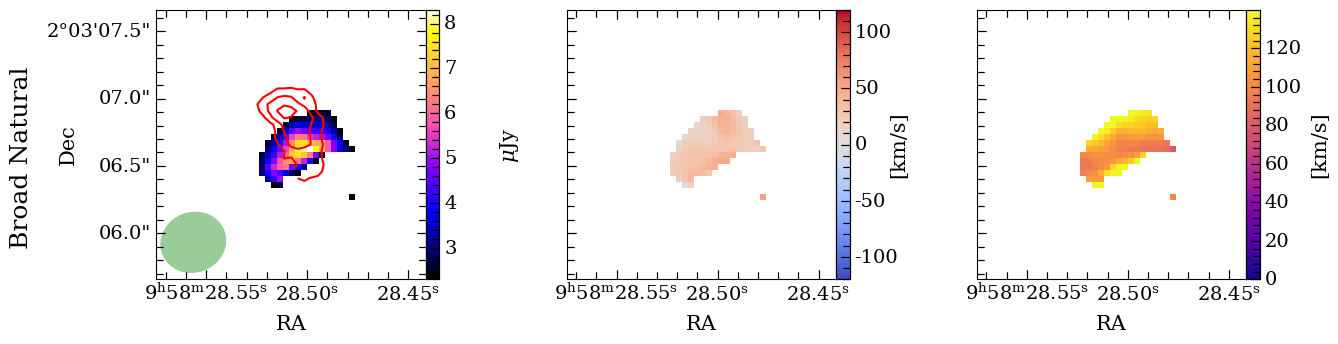}
    \includegraphics[width=\hsize]{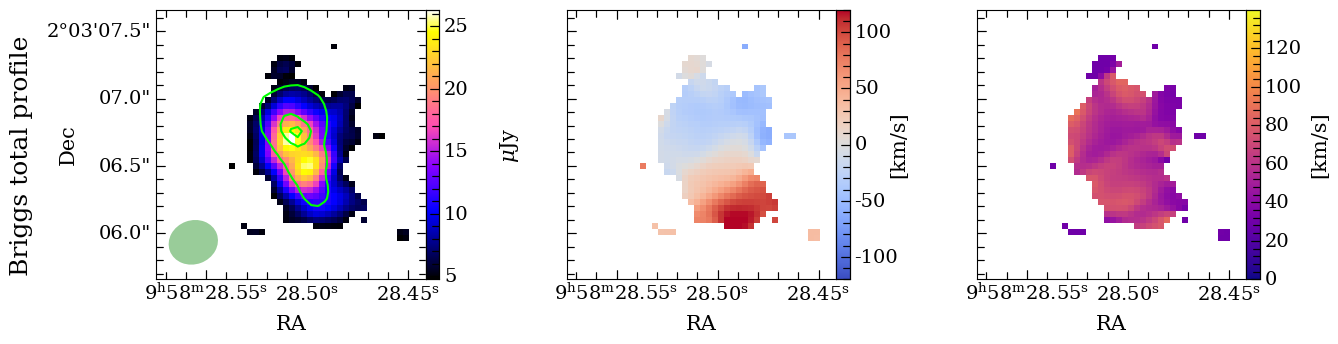}
    \caption{\cii\ moment maps derived from a spaxel-by-spaxel Gaussian fitting. From left to right we present the maps of the amplitude of the Gaussian, the velocity, and the velocity dispersion, respectively.
    From top to bottom, we show the narrow and the broad components from the natural cube and the total emission from the 1-component fit from the Briggs robust=0.5 cube. We note that no broad components were selected from the Briggs cube because they fall below the 3$\sigma$ threshold.
    In the left panels, we show the ALMA beam as a green ellipse and the narrow and broad \oiii\ flux from Fig. \ref{fig:moment_maps} as lime and red contours, respectively.}
    \label{fig:moment_maps_alma}
\end{figure*}

Here we present the results of the spaxel-by-spaxel analysis of the ALMA data targeting the \cii\ emission line.
In Fig. \ref{fig:cii_one_spaxel_fit} we show the best-fit results for the \cii\ emission line in one spaxel. 
We present the best-fit results by using one component to fit the emission line in the right panel, and by using 2 components in the left panel. 
The 1-component fit shows an excess around a velocity of $\sim 200$ \kms. These residuals completely vanish when the second component is added. The 2-component model is favored by the BIC test, with a $ \rm \Delta BIC = 14$. We perform this test on each spaxel and we end up selecting the broad component for 115 spaxels.

In Fig. \ref{fig:moment_maps_alma} we show the results for the spaxel-by-spaxel fitting of the \cii\ data. 
In the first two rows we show moment maps of the narrow and broad components from the natural weighting reduced cube, while in the panels in the bottom row, we show the moment maps from the datacube reduced with Briggs weighting, robust=0.5.
The cube reduced with natural weighting has a larger beam, but a higher S/N on a spaxel level with respect to the Briggs, robust=0.5 one. The higher S/N allows us to select a broad component, that we show in the central row.
This component is redshifted with respect to the host galaxy.
This redshifted broad component is what was already observed in circular apertures by \cite{Herrera-Camus:2021}, similarly elongated along the northwest direction.
In contrast to the \oiii\ broad component, it is redshifted with respect to the host galaxy ($\Delta v \sim 35$ \kms), but the projection is co-spatial with the \oiii\ broad (shown as the red contours).
The broad \cii\ component appears extended toward the filamentary structure (feature in the northwest at RA $\sim$ 9h58m28.47s, Dec $\sim$ 2d0.3m07.0s) that we observe in the narrow component natural flux map and in the map of the total profile.
The moment maps of the narrow component show an elongated \cii\ emission 
due to the presence of two separate spatial components, one peaking where the \oiii\ peaks (the north component), one at the location of the central component, where the integrated \cii\ map from Fig. \ref{fig:astrometry} peaks.
The smeared velocity field resembles that of a rotating galaxy, as already observed in other studies \citep{Herrera-Camus:2022, Parlanti:2023}. 

In the bottom panels of Fig. \ref{fig:moment_maps_alma} we show the maps of the narrow component obtained from the fitting of the Briggs cube. 
The Briggs robust = 0.5 cube unambiguously confirms the presence of two distinct clumps (northern and central components) in the amplitude map, through the higher spatial resolution. 
The peak \cii\ flux is in the location of the central obscured component. This is in agreement with the \cii\ being a tracer of star formation as the maximum luminosity arises from the regions with higher star formation (see also Sect.~\ref{sec:SED}).

\subsection{Dust continuum emission}

The ALMA observations also allow us to study the resolved dust continuum emission at $\sim 160~\mu$m in the galaxy rest frame.
While the optical and UV rest-frame emission allows us to probe directly the light emitted from the young stars, the FIR continuum emission traces the thermal radiation emitted from the dust that is heated by the stars. As the UV emission is absorbed by the dust, we can probe the obscured SF in the system by analyzing the dust emission.

\begin{figure}
    \centering
    \includegraphics[width=\hsize]{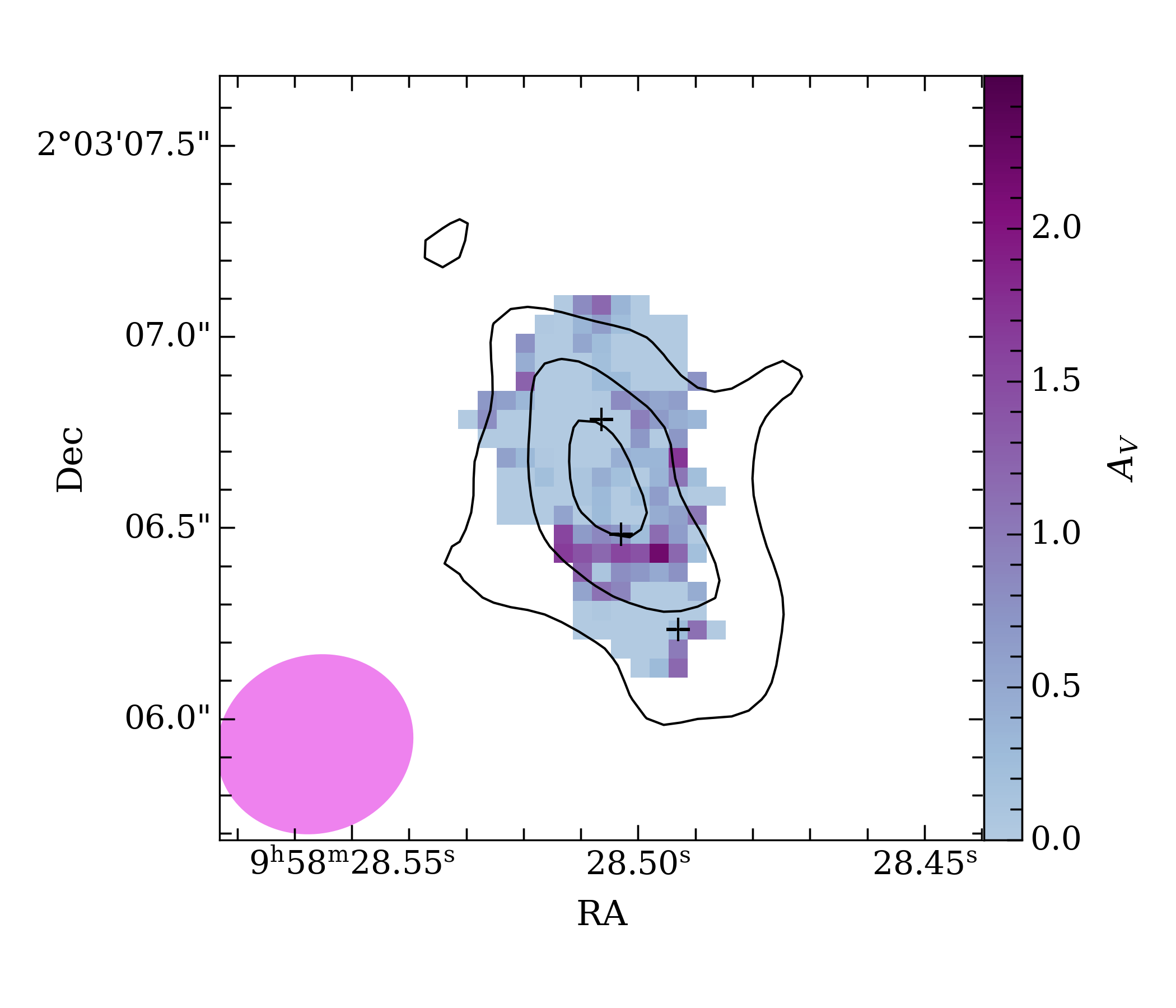}
    \caption{$A_V$ map from narrow component (same as in Fig.~\ref{fig:dust_sfr}) and dust continuum emission contours at 160~$\mu$m rest-frame.
    In black we report the dust continuum 3, 6, and 8$\sigma$ contours. The black plus signs represent the location of the three components identified from the \oiii\ channel maps and in optical continuum emission. The purple ellipse represents the beam size of the dust continuum emission.
    }
    \label{fig:dust_continuum}
\end{figure}

In Fig. \ref{fig:dust_continuum}, we report the dust emission contours at 3, 6 and 8 $\sigma$ on top of the $A_V$ map.
The dust emission morphology follows the elongation of the source, already seen in \oiii\ and \cii\ emission. The dust continuum peaks between the northern and the central components, where the $A_V$ measured through the Balmer decrement is 0 instead. The beam size ($0.46\arcsec\times0.42 \arcsec$) is larger than the separation between the two components.
The difference between the $A_V$ map and the dust continuum emission can be ascribed to a difference in the column density and the dust temperature across the field of view.
The total flux of the source at 160~$\mu$m is $S_{\rm 160\,\mu m} = 134 \pm 33~ \mu$Jy, comparable with the values reported in \citet{Herrera-Camus:2021} and \citet{Mitsuhashi:2023} with a similar resolution and \citet{Faisst:2020_dusttemp} with lower resolution.

The observations of the FIR continuum, together with the UV continuum and \ha\ emission line, allow us to compare the SFR computed with different methods.
Before the advent of JWST, the total SFR of high-$z$ galaxies, due to the lack of optical rest-frame emission observations, was estimated by summing the unobscured SFR estimated from UV continuum emission, and the obscured one estimated from the FIR continuum.
The three, low-resolution ($\sim$1\arcsec), dust continuum detections in ALMA bands 6, 7, and 8 of HZ4 allow us to perform a FIR SED fitting to estimate the obscured SFR. \citet{Faisst:2020_dusttemp} report $\rm SFR_{UV} = 29\pm4$~\sfr\ exploiting the UV HST observations, and through SED fitting they estimate a $\rm SFR_{IR} = 77_{-69}^{+104}$~\sfr.
Despite three FIR detections, the $\rm SFR_{IR}$ has uncertainties of $\sim 1$ dex, principally driven by the uncertainties in determining the dust temperature ($T_{D} =57.3^{+67.1}_{-16.6}$ K; \citealt{Faisst:2020_dusttemp}) due to the lack of observations near the IR peak, which can strongly constrain the temperature.

Thanks to the JWST observations, we were able to determine the galaxy SFR thanks to the observed \ha\ luminosity, corrected for the dust obscuration according to the Balmer decrement.
Assuming that the dust-corrected \ha\ luminosity is able to recover the entire intrinsic SFR ongoing in the system, we obtain that the total SFR of the system is $\rm SFR_{H\alpha} = 77^{+19}_{-16}$ \sfr.
Of this, $29\pm4$ \sfr\ is unobscured SF, and the remaining is obscured SF of $\rm SFR_{obscured} = 48^{+23}_{-20}$ \sfr.
By using the relation between the obscured SFR and $L_{\rm IR}$ by \citet{Kennicutt:2012} $\log{(\rm SFR/[M_\odot\, yr^{-1}]}) = \log(L_{\rm IR}/[\rm erg\, s^{-1}]) - 43.41$, we infer an IR luminosity of $\log(L_{\rm IR}/L_\odot) = 11.5 \pm 0.2$.
The IR luminosity obtained is comparable within the errors to the value obtained in other works from the SED fitting \citep{Faisst:2020_dusttemp,Fudamoto:2023}.

Then, we can now perform a SED fitting for the three FIR detections (we adopt the values from \citealt{Faisst:2020_dusttemp} for the band 6 and 8 fluxes and the value reported above for the band 7 detection), but constraining the $L_{\rm IR}$ from the obscured SFR inferred above.
We assumed that the IR and \ha\ trace SF over a similar timescale, a dust emissivity index of $\beta = 2$, and we performed a fit assuming a gray-body curve with a single temperature.
With these assumptions, we obtain a dust temperature of $T_D = 70\pm12$~K and a dust mass of $\log(M_D/M_\odot) = 6.62 \pm 0.17$. 
The dust temperature is in agreement with the values of \citet{Faisst:2020_dusttemp} and \citet{Fudamoto:2023}, but with smaller uncertainties.

\section{Discussion}

\label{dsec:discussions}

Based on previous observations, HZ4 was considered a typical high-z turbulent rotating disk. 
Our new JWST/NIRSpec IFS data instead reveal a clumpy structure, a different morphology depending on the tracer UV continuum emission, \oiii$\lambda$5007\AA and other rest-frame optical emission lines, \cii\ or dust continuum, with each clump having different SFHs and different properties, and an asymmetric velocity field.
These new observations hint at this galaxy being a merger between at least two galaxies with different velocities, instead of a rotating disk.

\begin{figure*}[ht!]
        \centering
    \includegraphics[width=\hsize]{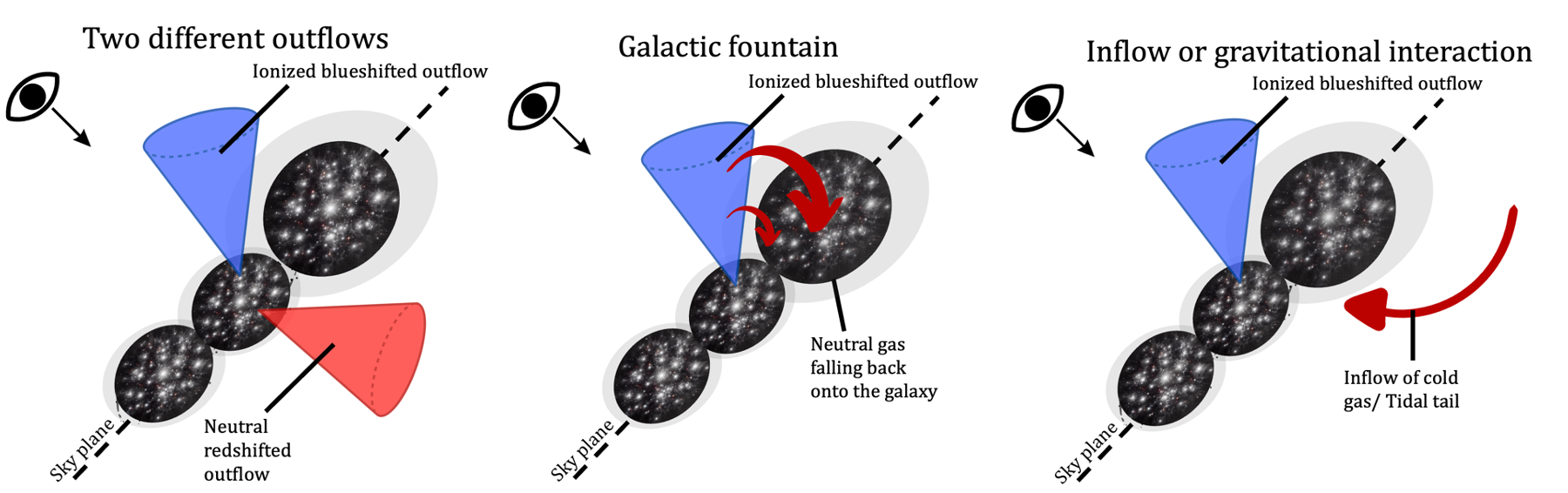}
    \caption{Cartoon illustrating three possible scenarios involving the presence of outflows for the explanation of the observations of the broad lines. In all panels, we draw in red the \cii\ noncircular components, and in blue the ionized gas emitting the broad rest-frame optical lines.
    In the left panel, we represent the multiphase outflow scenario in which the ionized broad line is mainly emitted by the approaching side of the outflow launched by the SF clump. The outflow traced by the \cii\ broad component is emitted by outflowing gas on the receding side.
    In the central panel, we show the galactic fountain scenario, in which the ionized gas is still tracing the outflow, but the \cii\ emission is tracing the gas that cannot escape from the potential well of the galaxy, then it cools down and falls back onto the galaxy. 
    The right panel instead depicts the scenario in which the redshifted \cii\ emission is tracing the inflow of cold gas from the CGM or IGM onto the galaxy, or is tracing gas stripped by the outer region of the northern and southern galaxies due to gravitational interaction.}
    \label{fig:scenarios}
\end{figure*}
 
The presence of additional broader components, statistically needed to reproduce the observed line profile, also implies that there are noncircular motions both in the rest-frame optical lines, which trace the ionized gas, and in the re-analyzed \cii\ emission line, tracing mainly the neutral gas and confirming the previous results by \citet{Herrera-Camus:2021}.
We discuss below different scenarios that can create the observed broad features (see  Fig. \ref{fig:scenarios}).

\begin{itemize}
    \item \textbf{Different outflows in different directions.}

This scenario is supported by the observation of a star forming and obscured clump, which thanks to the SF feedback is able to launch a blue-shifted, ionized gas outflow with the energetics that we observe.
On the other hand, the redshifted neutral outflow can be launched by the same region or other star-forming clumps (such as the \oiii\ bright northern clump), triggered by the merger event. The ionized and molecular outflows are almost co-spatial, but the \cii\ outflow extends in the northwest direction, while the ionized outflow extends in the northeast direction from what we assumed as the launching region. 
However, the $\Sigma_{SFR}$ is high enough to allow the launch of these outflows from the same regions.
We note that the observation of the redshifted outflow only in \cii\ does not exclude the possibility of it also hosting ionized gas. If the receding side of the outflow lies behind the disk with respect to our line of sight, that side of the ionized outflow could be obscured by the dust in the system in the optical emission lines observed by NIRSpec, but still be observed in \cii\ with ALMA. In that case, we would only be able to detect the blueshifted approaching side. 
We note that a similar behavior has been observed in \citet{Solimano:2024_ALMA} and \citet{Solimano:2024_JWST}, where a redshifted \cii\ plume is visible in ALMA data, and in the launching region there are signs of blueshifted outflow emission shown by JWST data.  
Additionally, observation of a local merger and starburst indicate a decoupling in the kinematics of ionized and neutral gas. This suggests that outflows, particularly in starburst-driven mergers, may have a complex geometry that cannot be modeled as simple conical outflows \citep{Shih:2010}.

\item \textbf{Galactic fountain.}

The redshifted \cii\ emission can also be explained as a galactic fountain. The ionized wind is not fast enough to escape the potential well of the galaxy as we will discuss in Sect. \ref{sec:outflow_strenght}, hence eventually the expelled material may cool down and then fall back onto the galaxy itself creating the galactic fountain \citep{Leroy:2015}. It is expected that the majority of the gas launched outward by SF-driven winds eventually will fall back as galaxy fountains \citep{Bregman:1980, Oppenheimer:2010, Brook:2012, Ubler:2014}. Given the low outflow velocity observed in comparison to the escape velocity, eventually this will happen, but it is unclear whether we are witnessing it now in these observations.
Unfortunately, the presence of a galactic fountain is hard to prove even in the local Universe and requires a detailed kinematic model of the galaxy and the outflow to prove or disprove its presence \citep[e.g.,][]{Yuan:2023}.

\item \textbf{\cii\ inflow or merger-induced flows and \oiii\ outflow.}
    The accretion of cold gas onto galaxies is one of the main mechanisms of galaxy growth and evolution. 
    Inflow signatures are hard to observe and study at high redshift, and are usually seen thanks to absorption features \citep{Rubin:2012, Bouche:2013, Herreracamus:2020}, or thanks to hints of radial motion via kinematic analysis \citep{ Arribas:2023, Genzel:2023, Ubler:2024_gn20}. 

If the most probable scenario is that the \oiii\ emission traces outflows,
the \cii\ emission instead can be associated with the inflow of cold atomic and molecular gas from the CGM and IGM.  As shown in Fig. \ref{fig:moment_maps_alma}, the orientation and location of the \cii\ broad component is aligned with the filamentary structure seen in the narrow component flux maps, both in the natural and in the Briggs cubes, which can resemble an accreting filament.

With this scenario the accreting gas must be already enriched in \cii, and thus cannot be pristine gas, hence the gas could originate from the galaxy, going back to the fountain scenario.

Although streams of inflowing gas may be present in the system, it is unlikely that they dominate the observed broad line profiles, which imply motions of $\sim 300$ \kms. While these velocities are typical of SF-driven outflows, they are too large for the inflow velocities that are expected to be a fraction of the virial velocity ($v_{\rm vir} = \sqrt{G M_{\rm DM}/r_{\rm vir}} \sim 180$ \kms;  \citealt{Goerdt:2015}).

Another possibility is that the broad wings observed may be explained as due to gravitational interaction (i.e., tidal tails) between these galaxies in the process of merging \cite[e.g.,][]{Baron:2024}.

\item \textbf{Shock due to the merging.}\\
The merger between the multiple components in the system can also trigger shocks, that increase the velocity dispersion of the gas.
In particular, during a major merger, tidal forces can compress and shocks the gas, triggering strong episode of star formation \citep{Pearson:2019}.
Observationally this can be observed as an increase in line ratios of low-ionization components and an increase in the velocity dispersion usually reaching in the local Universe $\sigma \sim 100 - 200$ \kms\ \citep{Rich:2011, Rich:2015}.
In our observations we find velocity dispersion of the broad component compatible with shocked regions in local galaxies (see Fig. \ref{fig:moment_maps}), however, we do not find line ratios consistent with the presence of shocks (see Fig. \ref{fig:BPT}), at least using the BPT and VO87 diagnostic diagrams.
The \oi\ emission line would be a better tracer of shocks but falls into the detector gap in our high-resolution dataset, and it is undetected in the R100 dataset. However, we can estimate a 3-$\sigma$ upper limit from the PRISM data of log(\oi/\ha) $< 10^{-1}$ for the total profile, which discourages the presence of strong shocks dominating in the ionization of the system \citep{Rich:2011}.
The current dataset does not allow us to put further constraints on the presence of the shock and their role in driving the observed increase in velocity dispersion of the system.

\end{itemize}

The present NIRSpec and ALMA observations describe a complex system formed by three merging galaxies, where likely gravitational interactions, inflowing gas, and outflows are present, with the latter explanation being the most probable in explaining the velocities implied by the broad component of the ionized emission line profiles. A better knowledge of its actual geometry would allow us to further constrain the properties of the system and analyze the relative role of these processes.



\section{Outflow properties}
\label{sec:outflow}

In this section, we assume that the broad emission lines seen in ALMA and JWST data are tracing a neutral and ionized outflow, and we derive its properties. With this assumption, we investigate the origin of the outflows and their launching mechanisms. Moreover, we present and discuss the outflow history.

\subsection{Ionized outflow properties}

Assuming that the measured broad lines are tracing the outflows, we first computed the integrated outflow properties for the ionized gas phase. By using the results from the modeling of the integrated spectrum reported in Table \ref{tab:r2700_fluxes}, we can compute the ionized gas mass from the flux of the \ha\ broad emission line component following \cite{Genzel:2011}, as

\begin{equation}
\label{eq:outflowmass}
    M_{\rm ion, out} = 3.2 \times 10^5 \left(\frac{\rm L_{\rm H\alpha, outflow}}{10^{40} ~\rm erg~s^{-1}}\right) \left( \frac{100~ \rm cm^{-3}}{n_e} \right) M_\odot,
\end{equation}
and the broad \oiii\ emission following \cite{Carniani:2023}, as

\begin{equation}
\label{eq:outflowmass_oiii}
    M_{\rm ion, out} = 0.8 \times 10^8 \left(\frac{\rm L_{\rm [OIII], outflow}}{10^{44} ~\rm erg~s^{-1}}\right) \left( \frac{500~ \rm cm^{-3}}{n_e} \right) \left( \frac{Z_\odot }{Z} \right) M_\odot,
\end{equation}

where $L_{\rm H\alpha, outflow}$ and  $L_{\rm [OIII], outflow}$ are the luminosity of the \ha\ and \oiii\ broad components corrected for the dust extinction, $n_e$ the outflow electron density and $Z$ the metallicity of the outflowing gas. We assume $n_e$ = 200 cm$^{-3}$, which is consistent with the value derived from the \siid\ ratio for the ISM of the galaxy with the integrated fit, and with density measured for outflows in high-$z$ galaxies \citep[e.g., ][]{Forster:2019}, and we adopted an outflow metallicity of $Z = 0.5 ~Z_\odot$ (see Sect. \ref{sec:analysis_R100}).
By using the value of \ha\ and \oiii\ fluxes reported in Table \ref{tab:r2700_fluxes}, we corrected for the outflow extinction of $A_V = 0.3_{-0.3}^{+0.9}$ and we estimate an ionized outflow mass of 
$M_{\rm ion, out} = (1.8 _{-0.5}^{+2.0})\times10^{7} M_\odot$ and 
$M_{\rm ion, out} = (2.5 _{-0.8}^{+4.5})\times10^{7} M_\odot$ for \ha\ and \oiii, respectively.

For the outflow velocity $v_{{\rm out}}$ we adopted the following relation from \citet{Rupke:2013} and \citet{Fiore:2017}:
\begin{equation}
\label{eq:outflow_velocity}
    v_{{\rm out}} = |\Delta v_{\rm narrow, broad}| + 2\sigma_{\rm out} = 389 \pm 12 {\rm~ km ~s^{-1}},
\end{equation}
where $\Delta v_{\rm narrow, broad}$ is the velocity shift between the peak of the broad and the narrow components and $\sigma_{\rm out}$ is the velocity dispersion of the broad component deconvolved for the instrumental broadening. With this assumption, we are accounting for the unknown geometry of the outflow, under the assumption that only the tail of the broad wings trace the outflowing gas directed in (or moving along) our line of sight, and thus its intrinsic velocity. 

The mass outflow rate is computed by assuming that the mass outflow rate is constant with time \citep[e.g.,][]{Lutz:2020}
\begin{equation}
\label{eq:outflowrate}
    \Dot{M}_{\rm out} = \frac{v_{{\rm out}}M_{{\rm out}}}{R_{\rm out}},
\end{equation}
where $R_{\rm out}$ is the extension of the galactic outflow. Based on the spaxel-by-spaxel fit we have inferred a flux-weighted radius of $R_{\rm out}\sim 2$ kpc (see Fig. \ref{fig:moment_maps}) that yields $\Dot{M}_{\rm out, H\alpha} = 2.3 ^{+2.7}_{-0.6}$ \sfr\ and $\Dot{M}_{\rm out, [OIII]} = 3.3 ^{+5.7}_{-1.1}$ \sfr.

\subsection{Neutral outflow properties}

With the new analysis of the ALMA data, we can also re-estimate the neutral atomic outflow properties under the assumption that the observed broad lines are tracing an outflow.
We computed the neutral atomic gas mass in the outflow by using the following equation:

\begin{equation}
    M_H = k_{[CII]}(T, n_H, Z) \times L_{[CII]}
\end{equation}
where $k_{[CII]}$ is a conversion factor that depends on the temperature ($T$), the neutral gas density ($n_H$), and the gas metallicity ($Z$), and $L_{[CII]}$ is the luminosity of the observed \cii\ outflow. 
Since the outflow temperature and density are unknown we conservatively compute the value of $k_{[CII]}$ assuming $T = 10^4$ K and $n_H=10^4$ cm$^{-3}$, which correspond to the maximal excitation case. With this assumption, the mass obtained represents a lower limit on the neutral gas mass traced by the \cii\ emission line.

Assuming an oxygen abundance of $\rm 12 + \log(O/H) = 8.34$ 
($Z=0.45~Z_{\odot}$, see Sect. \ref{sec:analysis_R100}), the case for maximal excitation gives $k_{[CII]}=4.88 ~M_\odot/L_\odot$ \citep{Herrera-Camus:2021}. The minimum neutral mass in the outflow is then $M_{\rm H, out} =2.69 \times 10^8 ~M_\sun$.
We compute the flux-weighted radius of the outflow from the map shown in Fig. \ref{fig:moment_maps_alma}, which gives $R_{\rm H, out} \sim 1.3$ kpc, and the outflow velocity computed as in Eq. \ref{eq:outflow_velocity} is $v_{\rm H,out} \sim 260$ \kms. Finally, we compute a lower limit on the mass outflow rate of $\Dot{M}_{\rm H, out} \sim 121$~\sfr. 
The value we obtain is higher than the lower limit reported in \citet{Herrera-Camus:2021} of $\sim$ 34 \sfr, but this is principally due to the difference in the $k_{[CII]}$ (\citealt{Herrera-Camus:2021} assumed $k_{[CII]} = 1.5~M_\odot/L_\odot$, which corresponds to maximal excitation assuming solar metallicity).

\begin{figure}
    \centering
    \includegraphics[width=\hsize]{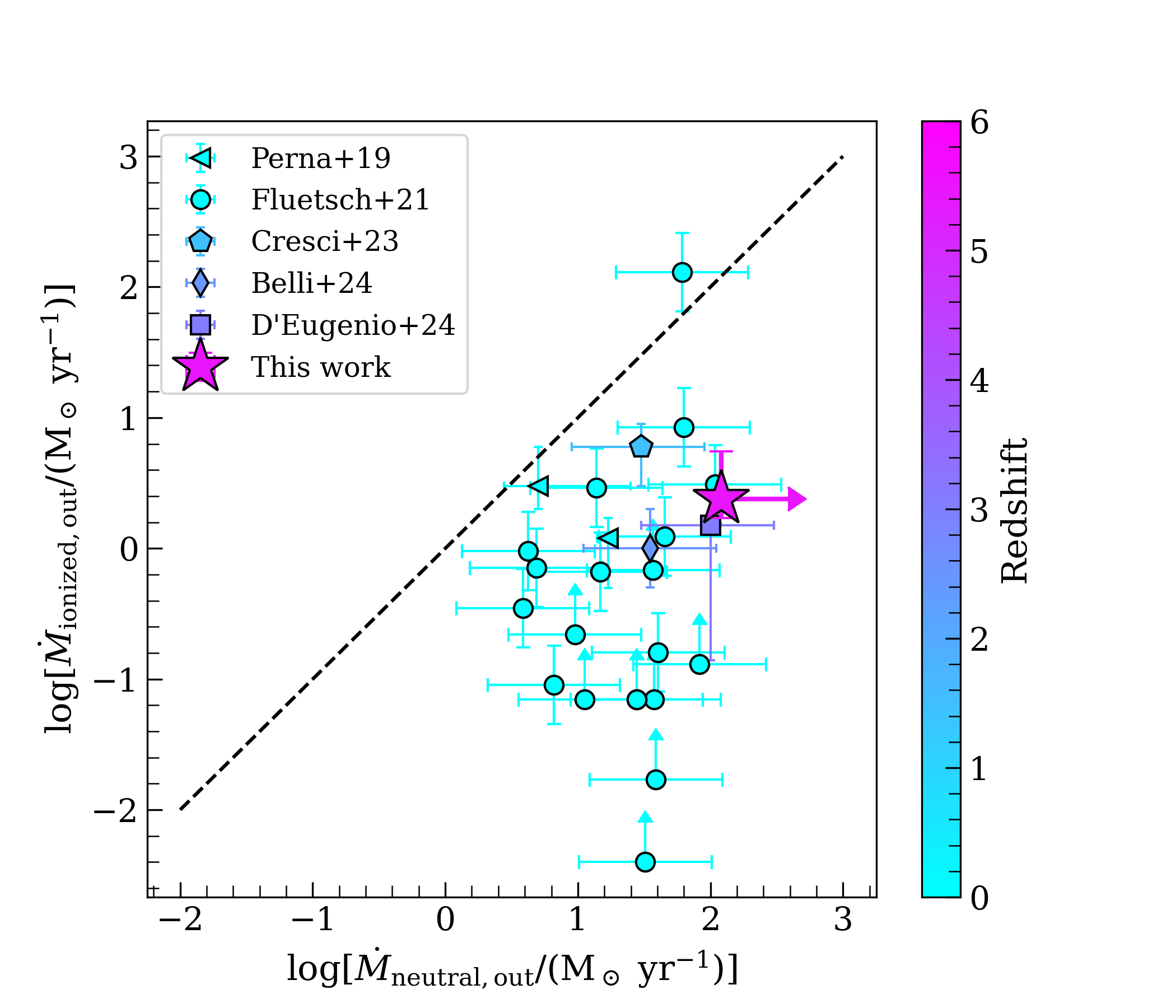}
    \caption{Comparison between the ionized and neutral mass outflow rate color-coded by redshift. The pink star marks the measurements for HZ4 from this work. The turquoise circles are the local ULIRGs studied in \citet{Fluetsch:2021}. The triangle, pentagon, diamond, and square display the results obtained for local and high-$z$ AGNs by \citet{Perna:2019, Cresci:2023, Belli:2023, Deugenio:2023}, respectively. The black dashed line represents the 1:1 relation. }
    \label{fig:moutrate_comparison}

\end{figure}

In HZ4, the neutral atomic mass outflow rate is more than one order of magnitude higher than the ionized mass outflow rate.
A similar behavior is observed both at low and high redshifts and for AGNs and SF galaxies \citep{Herreracamus:2020,Fluetsch:2021, Avery:2022, Cresci:2023, Deugenio:2023, Belli:2023}.
Analyzing a sample of local ULIRGs, \cite{Fluetsch:2021} shows that the amount of mass expelled by the galaxy in the ionized phase is usually negligible with respect to the neutral and the molecular phases, the latter of which accounts for the majority of the outflow mass. We compare our results with others from the literature in Fig.~\ref{fig:moutrate_comparison}, where we show that both AGNs and SF galaxies at low and high redshift lie below the 1:1 relation between the ionized mass outflow rate and the neutral one.

\subsection{Outflow strength and effect on the host galaxy}

\label{sec:outflow_strenght}

To understand whether these outflows are able to leave the galaxy, and hence remove gas from the system and enrich the CGM and the IGM, we estimate the escape velocity.
We followed the same approach as in \citet{Carniani:2023}. We modeled the galaxy gravitational potential as the sum of an exponential disk profile potential for the baryonic component with mass $ \rm M_\star + M_{\rm gas}$ (the dust mass, of $\sim 10^8$~\msun, is negligible; \citealt{Pozzi:2021}) and a Navarro-Frenk-White \citep[NFW;][]{Navarro:1996} profile for the potential of the dark matter halo. 

We considered as $ \rm M_\star$ the stellar mass of the whole system reported in Table \ref{tab:integratedproperties_bagpipes}. For the gas mass, we convert the \cii\ luminosity as $M_{\rm gas} = \alpha_{\rm [CII]} \times L_{\rm [CII]}$. We adopt $\alpha_{\rm [CII]} = 31 ~M_\odot/L_\odot$ \citep{Zanella:2018}.
To compute the mass of the dark matter halo, we adopted the stellar-to-halo mass relation from \citet{Moster:2013}, and we compute the concentration parameter for the NFW profile from the relation between the mass of the dark matter halo and the concentration estimated from \citet{Dutton:2014}.

We find that the velocity necessary for the gas to leave the galaxy and reach infinity is $v_{\rm esc} \sim 800$ \kms.
This velocity is higher than the estimated outflow velocity of the ionized and neutral gas, hence these outflows are leaving the launching site, but eventually, the gas may fall back on the galaxy.
We note that these calculations assume that the outflow moves ballistically (i.e., any other slowing mechanisms, such as the drag due to the gas encountered in the way, are not taken into account) and it is launched from the center of these idealized profiles.

To understand the impact that the outflow has on the removal of gas from the galaxy we estimate the mass loading factor $\rm \eta = \Dot{M}_{out}/SFR$.
For the integrated values we obtain $\eta_{\rm ionized} \sim 3\%$, while $\eta_{\rm neutral} \sim 150\%$. The ionized value is smaller than what found for other high-$z$, but less massive galaxies \citep[$\eta_{\rm ionized} \sim 200\%$ for galaxies with $\log(M_\star/M_\odot) \sim 7.5 - 8.5$;][]{Carniani:2023}. Both cosmological simulations and observations predict that the mass loading factor should decrease with increasing stellar mass for star-formation driven outflows, with mass loading factors reaching less than unity at a mass of around $\log(M_\star/M_\odot) \sim 10$,  consistent with what we observe \citep{Muratov:2015,Nelson:2019, Pandya:2021,Carniani:2023}.

\subsection{Outflow driver and launching region}

To investigate the outflow driver we first computed the energetics of the outflow.
We compute the kinetic and momentum rates as

\begin{equation}
    \Dot{E}_{\rm out} = \frac{1}{2} \Dot{M}_{\rm out} v_{\rm out}^2
\end{equation}
\begin{equation}
    \Dot{P}_{\rm out} = \Dot{M}_{\rm out} v_{\rm out}.
\end{equation}

For the ionized outflow, we obtain $\Dot{E}_{\rm out, H\alpha} = 1.16\times 10^{41}$ erg~s$^{-1}$, $\Dot{E}_{\rm out, [OIII]} = 1.57\times 10^{41}$ erg~s$^{-1}$, and
$\Dot{P}_{\rm out, H\alpha} = 5.9\times 10^{33}$ g cm s$^{-2}$, $\Dot{P}_{\rm out, [OIII]} = 8.1\times 10^{33}$ g cm s$^{-2}$ with uncertainties on the order of 0.5 dex.
Even though we cannot completely rule out the possible presence of an AGN (see Sect. \ref{sec:analysis_R2700}), the outflow velocity and the outflow energetics are small in comparison to what we expect from AGN-driven winds \citep{Fiore:2017}.
On the other hand, the outflow can be driven by the stellar feedback in a particular combination of winds from massive stars, and type 2 supernova (SN) explosions that release energy and momentum into the ISM \citep{Veilleux:2005} as explained below.
The SFR threshold for driving an outflow is considered to be around $\Sigma_{\rm SFR} \sim 1$ \sfr\ kpc$^{-2}$ \citep{Newman:2012, Davies:2019}. The star-forming obscured clump (HZ4-C) has a SFR of $\sim 20$ \sfr\ in a region of radius $\sim 0.9$ kpc, resulting in a $\Sigma_{\rm SFR} \sim 4$ \sfr\ kpc$^{-2}$  which is high enough to launch the ionized outflow.

If the outflow is driven by the stellar feedback from the burst in the central dusty region, we can compare the energy rate of the SN exploding in that region with the kinetic rate of the outflow.
We can compute the energy rate driven by SN explosions as $ \Dot{E}_{\rm SNII} = \epsilon E_{\rm SNII} R $, where $\epsilon$ is the supernova efficiency, that is the efficiency of transferring kinetic energy to the ISM, $E_{\rm SNII}$ is the total energy released in a SN explosion (10$^{51}$ erg), and \textit{R} is the supernova rate.
Assuming a SFR of $\sim 20$ \sfr, for a Kroupa IMF, the rate of SN explosions is related to the SFR as $R \sim 0.01 ~ \rm yr^{-1} \times {\rm SFR}$ / (\sfr) \citep{Mo:2010}.
If we assume a SN efficiency of $\epsilon \sim 0.1$,  we obtain  $ \Dot{E}_{\rm SNII} = 7.3 \times 10^{41}$ erg s$^{-1}$ which is higher than the outflow kinetic rate ($\Dot{E}_{\rm out} = 1.6 \times 10^{41}$ erg s$^{-1}$).
Following \citet{Heckman:2015}, the momentum rate injected by a starburst in the ISM is $\Dot{P}_{\rm SB} = 10^{33.7} \rm SFR$ / (\sfr) g cm s$^{-2}$. In the case of the central clump, we obtain $\Dot{P}_{\rm SB} \sim 10^{35} $ g cm s$^{-2}$.
Both the momentum and energy released by star formation are higher than those inferred for the outflow, thus the SF feedback in the central, obscured clump can drive the ionized outflow that we observe in the rest-frame optical emission lines.

Moreover, we observe that the outflow has higher $\Delta v$ and FWHM near the region of higher SFR (see Fig. \ref{fig:moment_maps} and \ref{fig:dust_sfr}), which corresponds to higher outflow velocity corrected for the unknown geometry. 
This is what we expect if that is the launching region and the outflow moves ballistically and slows down due to gravity, and possibly due to drag by gas encountered.

Since  the position of the peak of the outflow flux and the narrow \oiii\ flux do not match, we can infer that the outflow peak is not an artifact due to the higher S/N of the emission line, but rather a real higher flux of the outflow in that region.
The interpretation of these data is that the dust-obscured clump launches the ionized outflow which moves toward the north in the foreground of the galaxy with respect to our line of sight. A schematic representation of this interpretation is presented in the top-left panel of Fig. \ref{fig:scenarios}.

\subsection{Ionized outflow history}

\begin{figure}
    \centering
    \includegraphics[width=\hsize]{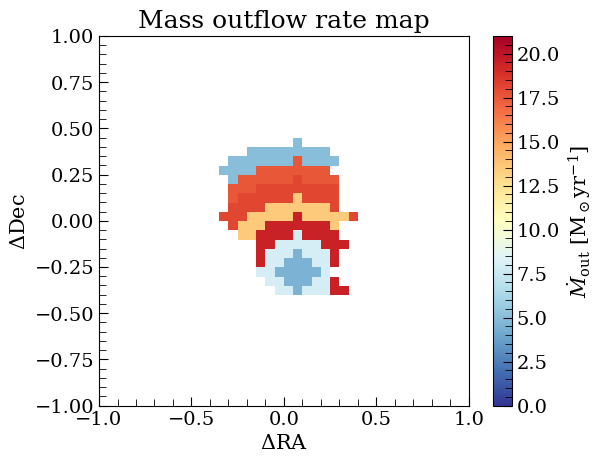}
    \caption{Ionized mass outflow rate map from the dust-corrected broad \ha\ emission for each radial shell.}
    \label{fig:outflow_shell}

\end{figure}

\begin{figure*}
    \centering
    \includegraphics[width=0.45\hsize]{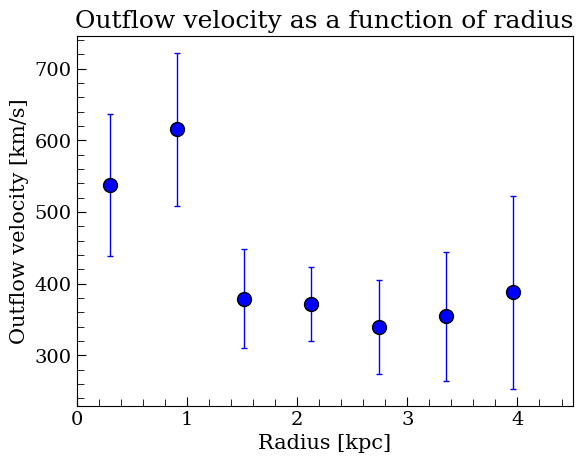}
    \includegraphics[width=0.45\hsize]{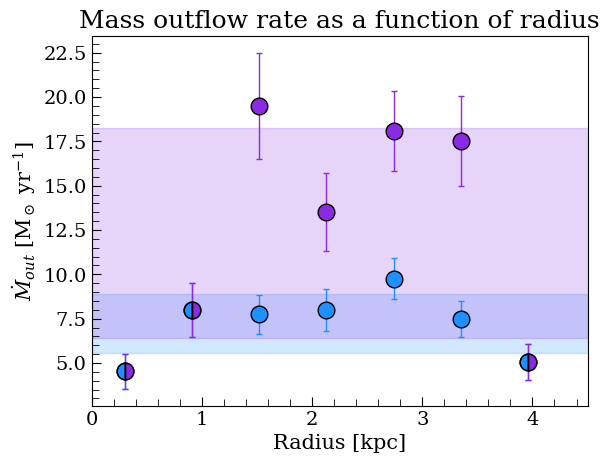}
    \includegraphics[width=0.44\hsize]{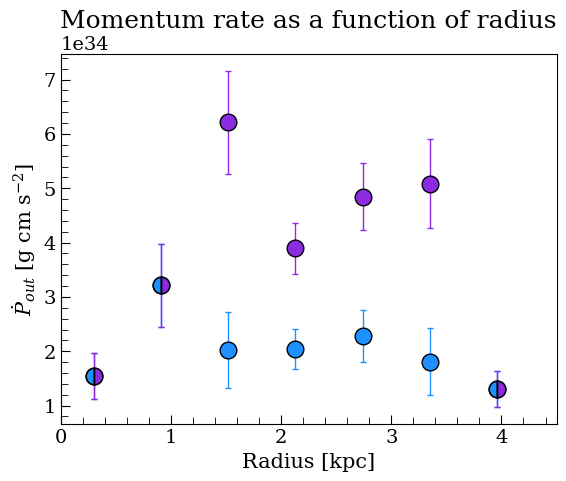}
    \includegraphics[width=0.45\hsize]{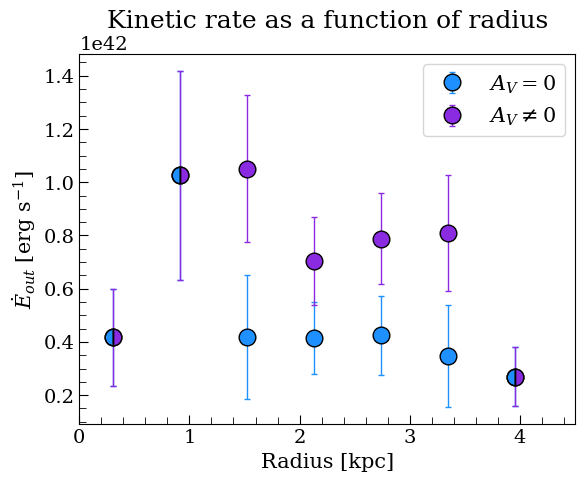}

    \caption{
    Outflow velocity (top left), mass outflow rate (top right), momentum rate (bottom left), and kinetic rate (bottom right) in each radial shell as a function of radius from the central component (see \ref{fig:outflow_shell}).
    Blue and purple points indicate whether the outflow mass was calculated by assuming $A_V = 0$ or the $A_V$ from the outflow Balmer decrement in each ring, respectively. The half-blue, half-purple points have measured $A_V = 0$.
    The purple and blue shaded regions in top-right panel show the 1$\sigma$ region around the mean outflow rate computed with $A_V\ne 0$ and $A_V = 0$, respectively.}
    \label{fig:outflow_radial}

\end{figure*}

Speculating that the broad component is an outflow, and assuming that the ionized outflow is being launched by the central, obscured, highly star-forming component, we extracted integrated spectra from 6 concentric annuli centered on that clump and spaced in radius by 0.1\arcsec, which corresponds to 609 pc at $z=5.54$. 
For each shell, we analyzed the spectra following the method reported in Sect.~\ref{sec:analysis_R2700}. 
Then we calculated the outflow properties as a function of radius from the launching region.
We computed the mass outflow rate for each shell following \citet{Lutz:2020} as
\begin{equation}
\label{eq:outflow}
    \Dot{M}_{\rm out, shell} = \frac{v_{{\rm out, shell}}M_{{\rm out, shell}}}{\Delta R}
\end{equation}
where $\Delta R = 609$ pc, being the size of each shell,  $v_{\rm out, shell}$ is the velocity of the outflow computed as in Eq. \ref{eq:outflow_velocity}, and $M_{\rm out, shell}$ is the mass in the shell calculated as using Eq. \ref{eq:outflowmass}. The outflow \ha\ luminosity was corrected for dust obscuration using the Balmer decrement computed in each shell.
In this case, the mass outflow rate represents the amount of gas passing through each projected shell. We show the mass outflow rate map in Fig. \ref{fig:outflow_shell}, in which each shell is color-coded by the mass outflow rate.

In Fig. \ref{fig:outflow_radial} we show the outflow velocity (computed using Eq. \ref{eq:outflow_velocity}), the mass outflow rate, the momentum rate, and the kinetic rate for each radial shell. 
We report all the quantities, assuming either no obscuration ($A_V = 0$, blue points) or the inferred obscuration based on the Balmer decrement in each shell ($A_V \neq 0$, purple points).
In the first two rings and the last one, the ratio between the broad \ha\ and \hb\ components is lower than the theoretical ratio of 2.86, so we assumed $A_V = 0$.

The observed outflow velocity is higher in the two rings closest to the launching region, then it decreases and stays almost constant up to 4 kpc.
This can also be seen in the outflow maps from the spaxel-by-spaxel fitting in Fig. \ref{fig:moment_maps} and \ref{fig:out_velocity_sfr}.
The mass outflow rate for each shell instead reaches its maximum between 1.4 and 3.5 kpc from the launching region.
The momentum and kinetic rate radial profiles show a steady increase with the radius up to 3.5 kpc in both quantities in the case in which we corrected for the dust attenuation of the outflow.

If we assume that each shell has moved at a constant velocity $v_{\rm out}$ in the plane of the sky since its launch, then the gas in each shell represents the outflow ejected at a time $\tau = v_{\rm out}/l$, where $l$ is the intrinsic, not projected distance from the launching point to where it is observed now.
With this assumption, we deduce that $\Dot{M}_{\rm out}$ was higher in the past since the mass outflow rate is higher at larger radii. 

\begin{figure}
    \centering
    \includegraphics[width=\hsize]{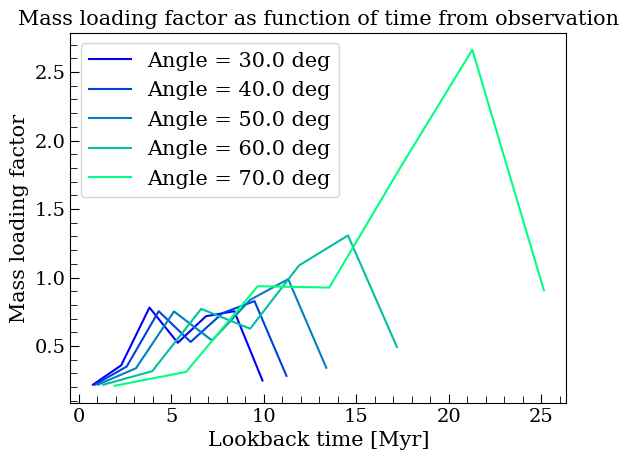}
    \caption{Mass loading factor of the ionized gas defined as $\Dot{M}_{\rm ion, out}$/SFR as a function of time for a range of assumed inclination angles.}
    \label{fig:mass_loading_factor_history}

\end{figure}

Assuming that the outflow has been launched at a time $\tau$ we can compare the outflow observed at a radius $R$, with the SFR in the central clump at the time $\tau= v_{\rm out}\times \cos(i)/R$, where $v_{\rm out}$ is the de-projected outflow velocity, $R$ is the distance along the plane of the sky and $i$ is the assumed inclination angle of the outflow with respect to the plane perpendicular to the line of sight.
We performed a SED fitting using \verb|Bagpipes| with the same parameters as reported in Sect. \ref{sec:SED}, but with finer SFH bins in the last 20 Myr to properly characterize the variation of the SFR in the last few Myrs when the outflow was launched. 
In Fig. \ref{fig:mass_loading_factor_history} we plot the mass loading factor $\eta$ as a function of lookback time, defined as $\eta = \Dot{M}_{\rm ion, out} (R)/ {\rm SFR}(\tau)$, where $\tau$ and $R$ are related by the equation reported above. For $v_{out}$ we assumed 450 \kms, which is the average outflow velocity across the different shells.
Varying the inclination angle $i$ between 30 and 70 degrees we can observe that the outflows must have been stronger in the past, reaching values of $\eta \gtrsim 1$ for $i>50$.
This analysis confirms what already observed by \citet{Herrera-Camus:2021} with the \cii\ neutral outflow, the mass loading factor of the outflow is low considering the entire galaxy ($\eta_{\rm ion} \sim 3\%$), but becomes significant when considering only the central component.

\section{Conclusions}
\label{sec:conclusion}

In this work, we have presented the JWST NIRSpec/IFS observations of the galaxy HZ4 at $z = 5.544$. This galaxy is the highest redshift galaxy for which we have evidence of a neutral outflow traced by broad \cii\ emission lines \citep{Herrera-Camus:2021}.
We have analyzed both the low- and high-spectral resolution datacubes from JWST NIRSpec IFS to study for the first time the rest-frame UV and optical spectrum of this source. Moreover, we combined our results with a re-analysis of the \cii\ ALMA data to constrain the multiphase properties of this galaxy.

In particular, our main results are:
\begin{itemize}

    \item The JWST high spatial resolution observations reveal that the galaxy is not a regular rotating disk as inferred from ALMA lower resolution observations, but rather it is formed by three different components. In particular, the different components have different properties, such as stellar ages and star formation histories.

    \item  We observe a spatially resolved broad component that we speculate it is tracing an ionized outflow, which is co-spatial with the broad component seen in \cii\ in our analysis and also in previous works.
    If interpreted as an outflow this would be the highest redshift case in which we are able to probe probe the SF-driven multiphase outflow.

    \item Assuming that the broad component is tracing an ionized outflow, this component is extended $\sim 4$ kpc from the launching site that we speculate is the highly star-forming region. With this assumption we constructed the ionized outflow history by analyzing the outflow properties in radial shells from the launching site. We observe that the velocity of the outflow decreases with increasing radius, while the mass outflow rate increases. We also computed the mass loading factor history of the outflow and found that the outflow was stronger in the past, reaching values of $\eta \sim 1 - 2$ at 5-20 Myrs, varying with the outflow inclination.

    \item If both the broad components in \oiii\ and \cii\ are outflows, we can estimate the multiphase outflow properties,  finding that in this $z\sim5.5$ source the mass of the SF-driven outflow is dominated by the colder gas phases, in agreement with other low-$z$ studies and a handful of studies up to $z \sim 3$.

\end{itemize}

The synergy between ALMA and JWST observations has allowed us to probe the multiphase ISM properties up to $z\sim5.5$.
These observations highlight also that high-resolution data are crucial for interpreting such complex systems. 
However, despite the large amount of observations for this source, it remains challenging to fully understand its geometry and pinpoint the origin of the broad components observed in both ALMA and JWST datasets.
Further observations and larger samples are needed to better assess the role of mergers and outflows in shaping galaxy properties in the early Universe.


\begin{acknowledgements}
SC, GV, and SZ acknowledge support from the European Union (ERC, WINGS,101040227).
SA and MP acknowledge grant PID2021-127718NB-I00 funded by the Spanish Ministry of Science and Innovation/State Agency of Research (MICIN/AEI/ 10.13039/501100011033).
RM and FDE acknowledge support by the Science and Technology Facilities Council (STFC), from the ERC Advanced Grant 695671 "QUENCH", 
RM acknowledges support by funding from a research professorship from the Royal Society.
FDE acknowledges support by the Science by the UKRI Frontier Research grant RISEandFALL.
H{\"U} gratefully acknowledges support by the Isaac Newton Trust and by the Kavli Foundation through a Newton-Kavli Junior Fellowship.
GC acknowledges the support of the INAF Large Grant 2022 "The metal circle: a new sharp view of the baryon
cycle up to Cosmic Dawn with the latest generation IFU facilities"
AJB and GCJ acknowledge funding from the "FirstGalaxies" Advanced Grant from the European Research Council (ERC) under the European Union’s Horizon 2020 research and innovation programme (Grant agreement No. 789056). 
IL acknowledges support from grant PRIN-MUR 2020ACSP5K\_002 financed by European Union - Next Generation EU.\\ This paper makes use of the following ALMA data: ADS/JAO.ALMA\#2018.1.01605.S, ADS/JAO.ALMA\#2012.1.00523.S,  ADS/JAO.ALMA\#2017.1.00428.L. \\ALMA is a partnership of ESO (representing its member states), NSF (USA), and NINS (Japan), together with NRC (Canada), MOST and ASIAA (Taiwan), and KASI (Republic of Korea), in cooperation with the Republic of Chile. The Joint ALMA Observatory is operated by ESO, AUI/NRAO and NAOJ.
\end{acknowledgements}
\bibliographystyle{aa}
\bibliography{aa}

\begin{appendix} 
 \section{Other sources in the FOV}
\label{appendix:foregroundsource}

\begin{figure}[h!]
    \centering
    \includegraphics[width=\hsize]{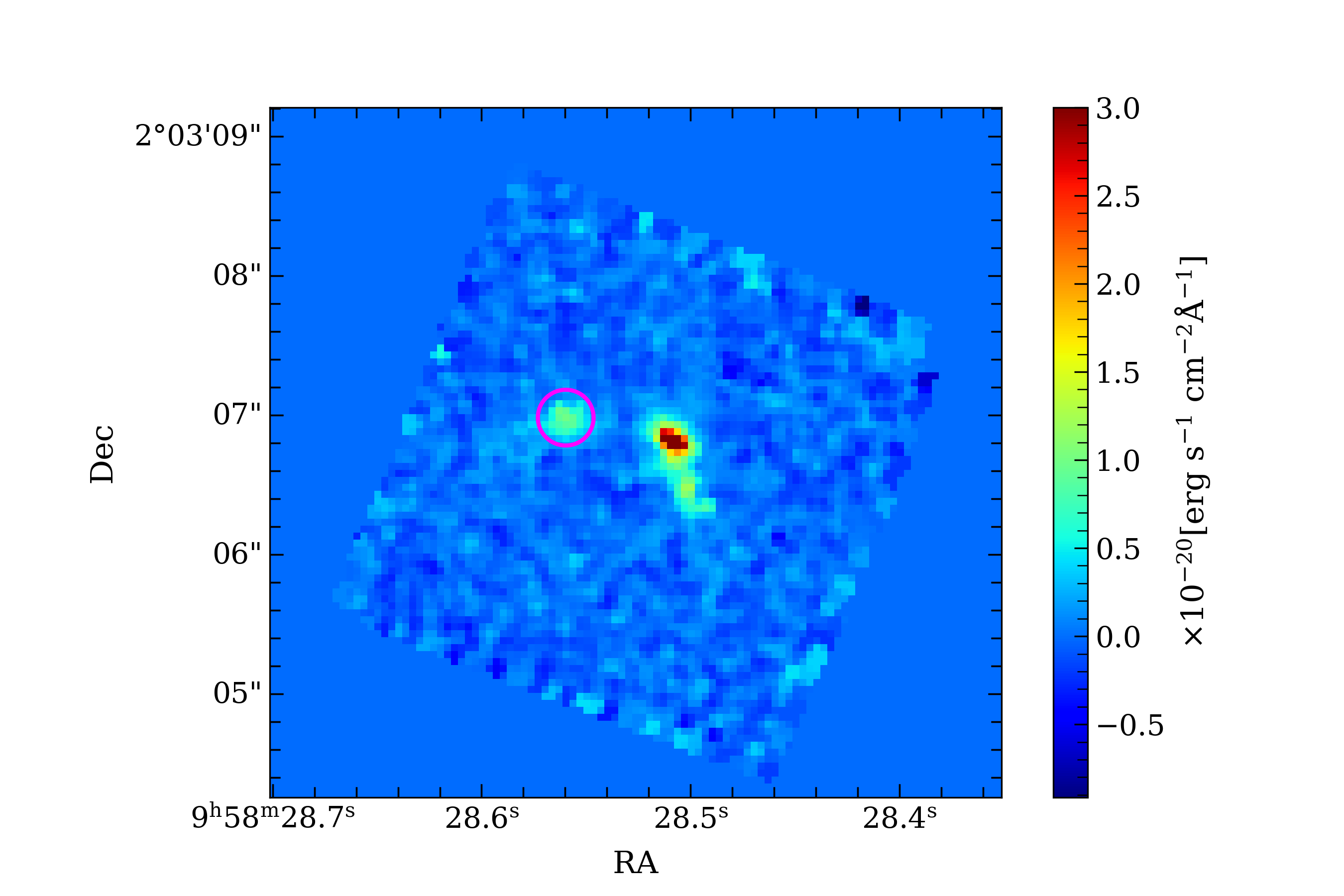}
    \includegraphics[width=\hsize]{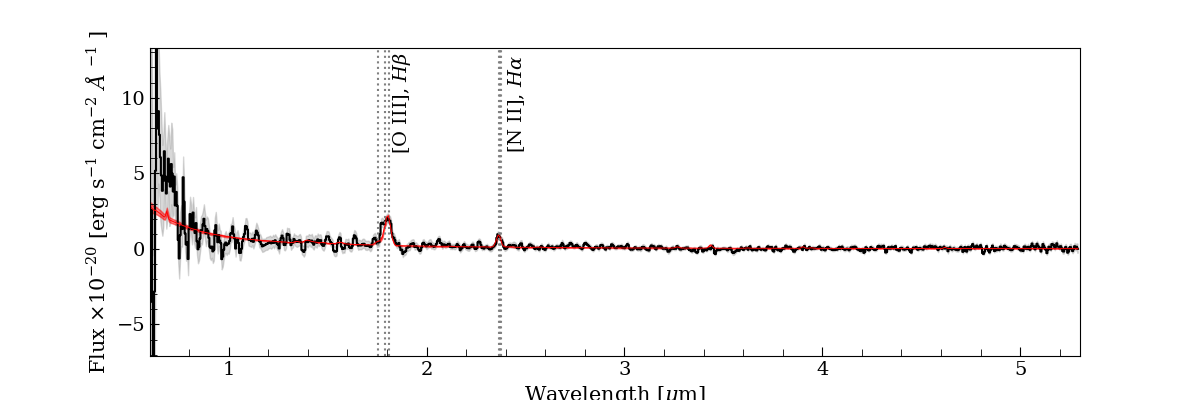}

    \caption{
    Top panel: R100 cube integrated between 1.77 and 1.84 $\mu$m. 
    Bottom panel: In black we plot the spectrum extracted from the area highlighted as the magenta circle in the panel above. In red we show the best-fitting result from the Bagpipes fitting.
    }
    \label{fig:extra_galaxy}
\end{figure}

While analyzing the R100 cube we detected another source in the field of view, which is distant $\sim$0.8\arcsec\ from the northern bright clump of HZ4. The source is located at RA = 09h58m28.55s, Dec = 02d03m06.69s, and it is undetected in HST or ALMA. The continuum emission of the source is negligible but is visible only when integrating between 1.77 and 1.84 $\mu$m as shown in the top panel of Fig. \ref{fig:extra_galaxy}.
In the bottom panel of Fig. \ref{fig:extra_galaxy} we show the spectrum of the source extracted from a circular aperture of radius 0.2\arcsec (4 spaxels) highlighted as the pink circle in the map.
The spectra show two bright emission lines, that we identify as the \oiii$\lambda\lambda$5007,4959\AA, \hb\ complex and the \ha\ and the \nii$\lambda\lambda$6548,6584\AA\ complex, which are unresolved in the prism spectrum.
We performed a fit of the emission lines by using the same procedure as reported in \ref{sec:analysis_R2700} and implying a single Gaussian component for each emission line. This places the galaxy at a redshift of $z = 2.607 \pm 0.003$.

We also estimate the galaxy properties by fitting the integrated spectrum with \verb|Bagpipes|.

The best-fit \verb|Bagpipes| spectrum is shown in the bottom panel of Fig. \ref{fig:extra_galaxy} as the red line. The galaxy has a stellar mass of $\log{(M_\star/M_{\odot})} = 7.2 \pm 0.1$.

\section{Outflow velocity}
\label{appendix:outflowvelocity}
In this Sect. we show the outflow velocity map computed as Eq. \ref{eq:outflow_velocity} for each spaxel and we compare it with the $\Sigma_{SFR}$ map (see also Fig. \ref{fig:dust_sfr}).

\begin{figure}[h!]
    \centering
    \includegraphics[width=\hsize]{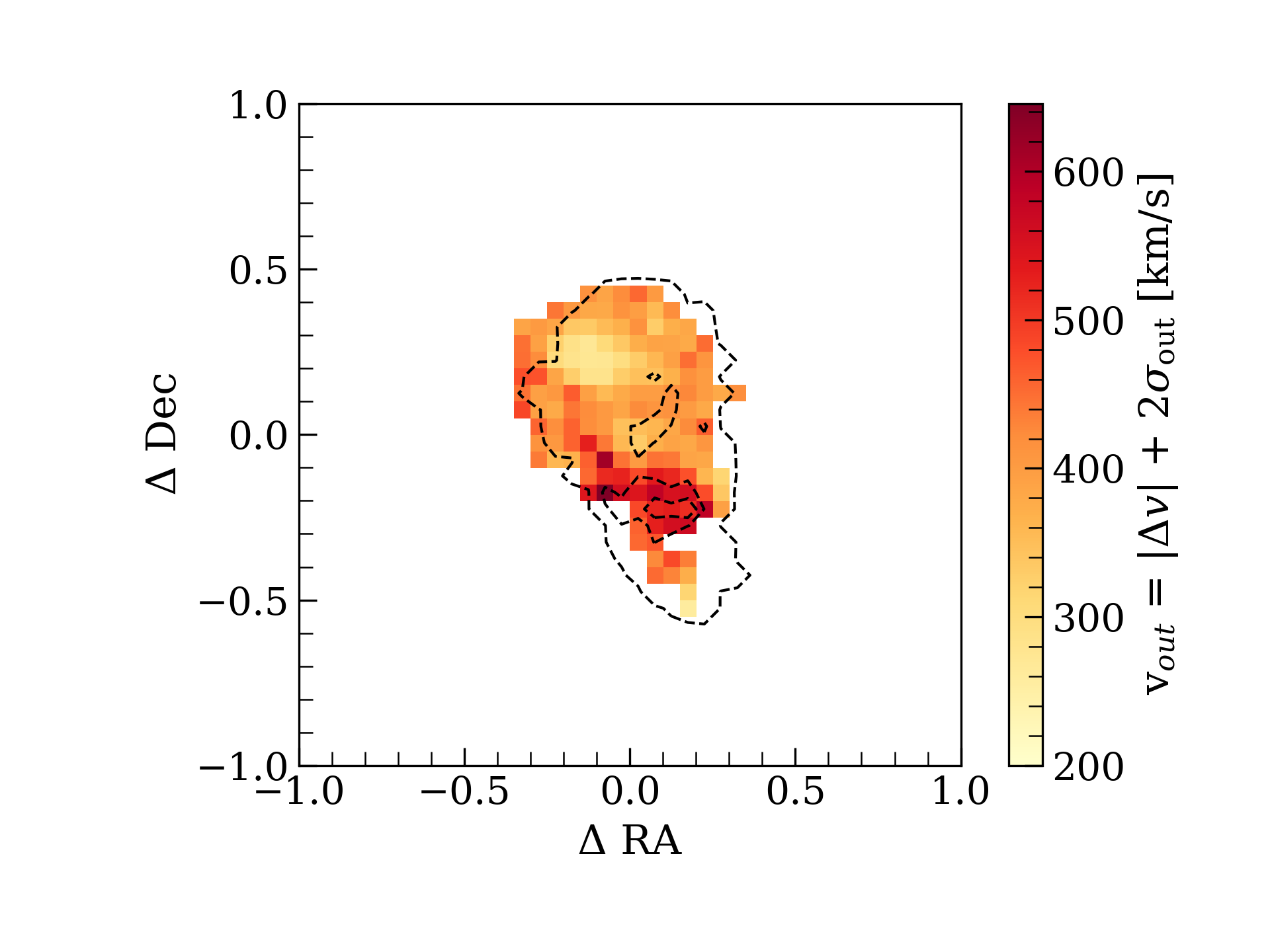}
    \caption{Outflow velocity map defined as $v_{\rm out}= |\Delta v_{\rm narrow, broad}| + 2\sigma_{\rm out}$ (see also Sect. \ref{sec:outflow}). We show as black dashed contours the $\Sigma_{\rm SFR}$ map contours (see also Fig. \ref{fig:dust_sfr}).}
    \label{fig:out_velocity_sfr}
\end{figure}

\newpage

\section{Outflow history}

\begin{figure*}[h!]
        \centering
    \includegraphics[width=\hsize]{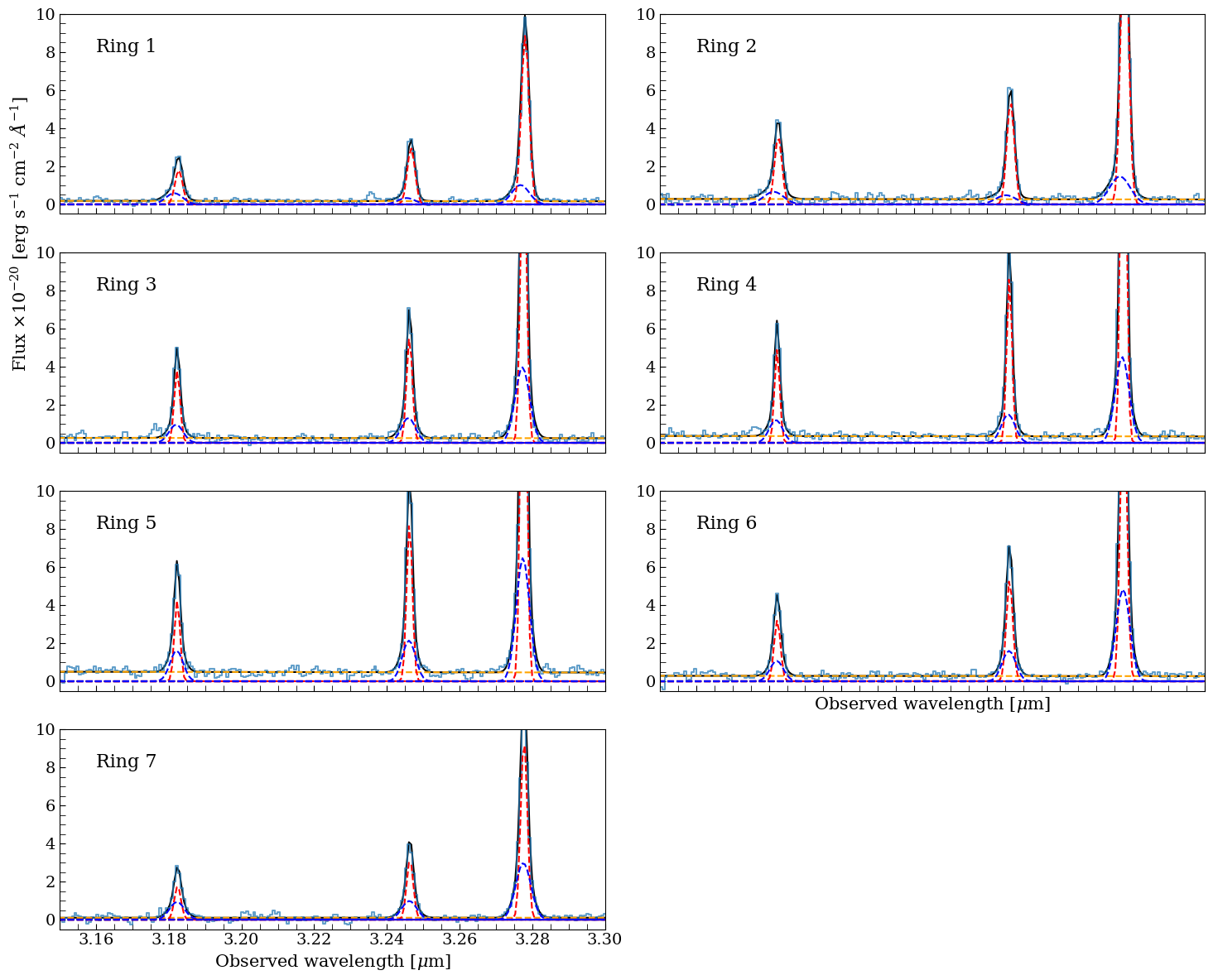}
    \caption{Zoom in on the \oiii\ and \hb\ complex of each ring. Red and blue dashed lines are the narrow and outflow best-fit models. Please note that the displayed range on the $y$ axis is the same for all the panels.}
    \label{fig:spectrum_each_ring_outflow}
\end{figure*}

In this Sect. we show the individual fit of each spectrum extracted from each ring for the analysis of the outflow history in Fig. \ref{fig:spectrum_each_ring_outflow}. Ring 1 represents the closer one to the launching site, Ring 7 is the farther. We show the \oiii, \hb\ complex.

\end{appendix} 
\end{document}